# Unveiling the Lithium-Ion Transport Mechanism in $Li_2ZrCl_6$ Solid-State Electrolyte *via* Deep Learning-Accelerated Molecular Dynamics Simulations

Hanzeng Guo[1,2], Volodymyr Koverga[1,2], Selva Chandrasekaran Selvaraj[1,2] and Anh T. Ngo[1,2]*

[1] Department of Chemical Engineering, University of Illinois Chicago, Chicago, IL 60608, United States

[2] Materials Science Division, Argonne National Laboratory, Lemont, IL 60439, United States

*Corresponding author's email: *anhngo@uic.edu*

## Abstract

Lithium zirconium chlorides (LZCs) present a promising class of cost-effective solid electrolyte for next-generation all-solid-state batteries. The unique crystal structure of LZCs plays a crucial role in facilitating lithium-ion mobility, which further affects its electrochemical performance. To understand the underlying mechanism governing ion transport, we employed deep learning-accelerated molecular dynamics simulation on $Li_2ZrCl_6$ (trigonal $\alpha$- and monoclinic $\beta$-LZC), focusing specifically on the zirconium coordination environment. Our results reveal that disordered $\alpha$-LZC exhibits the highest ionic conductivity, while $\beta$-LZC demonstrates significantly lower conductivity, closely aligning with experimental findings. The study confirms that across all phases, lithium migration proceeds via site-to-site hopping mechanism, where variations in site residence times critically impact the overall ionic conductivity. In $\alpha$-LZCs, lithium ions prefer to anisotropically diffuse across interlayers as the result of lower energy barrier, driven primarily by collective diffusion. In contrast, lithium ions in $\beta$-LZC primarily isotropically diffuse within intralayer, hindered by higher energy barriers and determined by individual diffusion. The variation in $ZrCl_6^{2-}$ octahedral unit softening, induced by the specific layered arrangement of zirconium atoms, emerges as a critical determinant of the energy barriers across the LZC phases. These atomic-scale insights into the transport processes provide valuable guidance for the rational design and optimization of LZCs-based electrolytes, accelerating their practical application in advanced energy storage technologies.

## Introduction

All-solid-state batteries (ASSBs) have attracted significant attention as next-generation energy storage devices with high energy density and excellent thermal and electrochemical stability [1-3]. For traditional lithium-ion batteries with liquid electrolyte, despite the well-established manufacturing process and low production cost, it still poses a significant security risk. Most liquid electrolytes are composed of flammable, usually carbonate-based, organic solvents, which can lead to an explosion in the case of internal short circuits, overheating, or mechanical damage. To address these concerns, ASSBs employ non-flammable inorganic solid materials instead of organic liquid electrolytes, offering higher safety and improved long-term lifetime [4-**Error! Reference source not found.Error! Reference source not found.**, including sulfide- [7], oxide- [8], and halide-based [9] systems. Interestingly, among the various families of solid-state electrolytes, halide-based systems have received considerable attention due to their essentially enhanced electrochemical and mechanical characteristics [10,11]. In particular, chloride-based solid electrolytes exhibit strong compatibility with high-voltage cathodes, positioning them as promising candidates for next-generation ASSBs and central focus of ongoing research efforts [12,13].

Nevertheless, apart from the numerous attractive features, for most chloride-based electrolytes reported to date, the high cost of raw materials remains a significant barrier to effective industrial application. To address this, $Li_2ZrCl_6$ (LZC) has emerged as one of the most economically viable candidates among chloride-based solid electrolytes owing to higher natural abundance and affordability of zirconium [14-23]. As previous experimental studies mentioned [14-21], the crystal structure of LZC is highly sensitive on its synthesis method. For example, under only ball milling and low temperature treatment, LZC adopts a hexagonal close-packed trigonal structure with $P$-3$m$1 space group, similar to the structures of $Li_3YCl_6$ and $Li_3ErCl_6$ [14]. In contrast, high-temperature treatment induces a phase transition to a monoclinic structure with cubic close-packing (space group $C2/m$), analogous to $Li_3InCl_6$ [19]. These two phases are commonly referred to as $\alpha$-LZC and $\beta$-LZC, respectively. It is also important to note that previous studies on $Li_3YCl_6$ and $Li_3ErCl_6$ halides electrolytes have demonstrated that site disorder in the $P$-3$m$1 space group can essentially influence ionic conductivity [24-26]. A similar phenomenon is observed in $\alpha$-LZC, where zirconium, which participate in the formation of a structural backbone, may also occupy multiple crystallographic site combinations, giving two variants of the same structure – ordered and disordered $\alpha$-LZC [17]. Specifically, the ordered $\alpha$-LZC exhibits a more closely packed multiple crystallographic layers, while in disordered form zirconium atoms are characterized by a loose arrangement with atoms distributed predominantly in a single layer.

The previously reported ionic conductivities of $\beta$-LZC and $\alpha$-LZC, *i.e.*, ordered and disordered, fall in range around 5.70-7.10 × $10^{-6}$ S $cm^{-1}$and 0.98-8.08 × $10^{-4}$ S $cm^{-1}$, respectively [14-21]. Notably, the conductivity of $\alpha$-LZC can vary up to a factor of two to three depending on synthesis parameters such as rotation speed and ball-to-powder mass ratio during the mechanochemical preparation [20]. Although the overall ionic conductivity of LZC remains lower than that of other chloride-based electrolytes containing $In^{-}$, $Y^{-}$, or $Er^{-}$ [26-28], its transport properties can be significantly improved through substitutional doping strategies. The prominent example of implementation of this approach is equivalent substitution with $Fe^{3+}$, where it was found that the ionic conductivity of $Li^{+}$ can be enhanced up 1 × $10^{-3}$ S $cm^{-1}$ [17]. Similarly, Mn doping can be used to potentially improve the first cycle performance during constant current charge and discharge of LZC as it was reported for halide-based electrolytes [29]. In another doping approach partial substitution of the chlorine with fluorine atom not only doubles the ionic conductivity compared to original LZC, but also significantly improves the stability under high pressure conditions [30]. Under these circumstances, LZC stands out with great potential to balance suitable manufacturing costs, large-scale production and high ionic conductivity, thus making further investigation of its ion transport mechanism particularly important.

However, experimental studies on LZC have yet to fully uncover the underlying physical and chemical transport mechanisms governing lithium motion which is the key factor determining the overall battery performance. To date, most theoretical studies of LZC have mainly focused on Density Functional Theory (DFT) calculations and *ab initio* molecular dynamics (AIMD) simulations of either $\alpha$-LZC or $\beta$-LZC [16,22,31,32]. Although DFT-based atomistic simulations accurately predict structural, electrochemical, and transport properties, their small-scale limitation hinders the understanding of lithium-ion transport-driven electrochemical behavior in real systems requiring long-time simulations. Additionally, for diffusion processes at higher temperatures, the finite-size effects inherent to small simulation cells may lead to an overestimation of the diffusion coefficient due to the insufficient statistical sampling [33][36]. To address these limitations, molecular dynamics (MD) simulations enhanced by machine learning (ML) techniques are increasingly being used [37-[39]. Among various ML-based methods to produce interatomic potentials, including Gaussian approximate potentials [40], artificial neural networks [41,42], atomic cluster expansion [43] and kernel-based methods [44], deep learning potentials (DLP) have gained particular attention due to their ability to generate highly accurate potential models based on first-principles data [45,46]. This approach enables simulation of thousands of atoms over extended time scale, therefore, bridging the gap between quantum accuracy and classical MD efficiency.

In this work, we adopted an integrated approach that combines AIMD, neural network-driven ML methods and classical MD simulations, commonly referred to deep learning-accelerated molecular dynamics (DLMD) simulations, to investigate the structural properties and transport properties of $Li_2ZrCl_6$. This DLMD achieves AIMD-level accuracy while enabling simulations at larger scales in terms of system size and simulation time. Using DLMD, we determined the ionic conductivities of ordered $\alpha$-LZC, disordered $\alpha$-LZC, and $\beta$-LZC in excellent agreement with experimental measurements. Furthermore, our analysis reveals that lithium diffusion in $\alpha$-LZCs is predominantly anisotropic with a strong preference for interlayer motion, whereas $\beta$-LZC exhibits nearly isotropic diffusion dominated by intralayer migration. Detailed analysis using the lithium probability density function and the van Hove correlation function indicates that lithium diffusion occurs through a site-to-site hopping mechanism, and variations in residence time at different structural sites are key factors driving the difference in ionic conductivity across the studied systems. Further analysis of the local structural organization confirms that variations in ion channel energy barriers, arising from softening of $ZrCl_6^{2-}$ octahedra induced by specific layered arrangements of zirconium atoms, explain the enhanced ionic conductivity observed in disordered $\alpha$-LZC compared to other studied phases. This comprehensive DLMD framework not only deepens our understanding of lithium transport but also establishes a robust methodology for guiding the rational design of next-generation energy storage devices.

With this, the manuscript is organized as follows: the *Simulation Methodology* section outlines the computational protocol to assemble DLP model and the details of the analysis of the obtained trajectories, the *Results and Discussion* section covers a detailed investigation of the lithium transport mechanism in LZC combining mean-squared displacement, van Hove correlation, residence time, spatial and radial distribution function analysis, potential of mean force, partial density of states and crystal-orbital Hamilton populations to reveal diffusion behaviors across different phases. Finally, the *Conclusions* section summarizes the key findings and discusses their implications.

**Simulation Methodology**

In this study, a hybrid deep learning-accelerated molecular dynamics (DLMD) approach was employed to accurately model lithium-ion transport in $Li_2ZrCl_6$ (LZC) electrolyte. The stepwise algorithm (**Figure 1**) integrates Density Functional Theory (DFT) and *ab initio* molecular dynamics (AIMD) simulations to generate high-quality reference data, which is then used to train deep learning potential (DLP) models. The trained DLP models possess the ability to capture the interatomic interactions with quantum accuracy while enabling simulations at significantly larger spatial and temporal scales compared to AIMD. The resulting DLMD simulations provide insights into the local structural organization and diffusion behavior, thereby elucidating the origins of the differences in ionic conductivity and activation energy across different LZC phases.

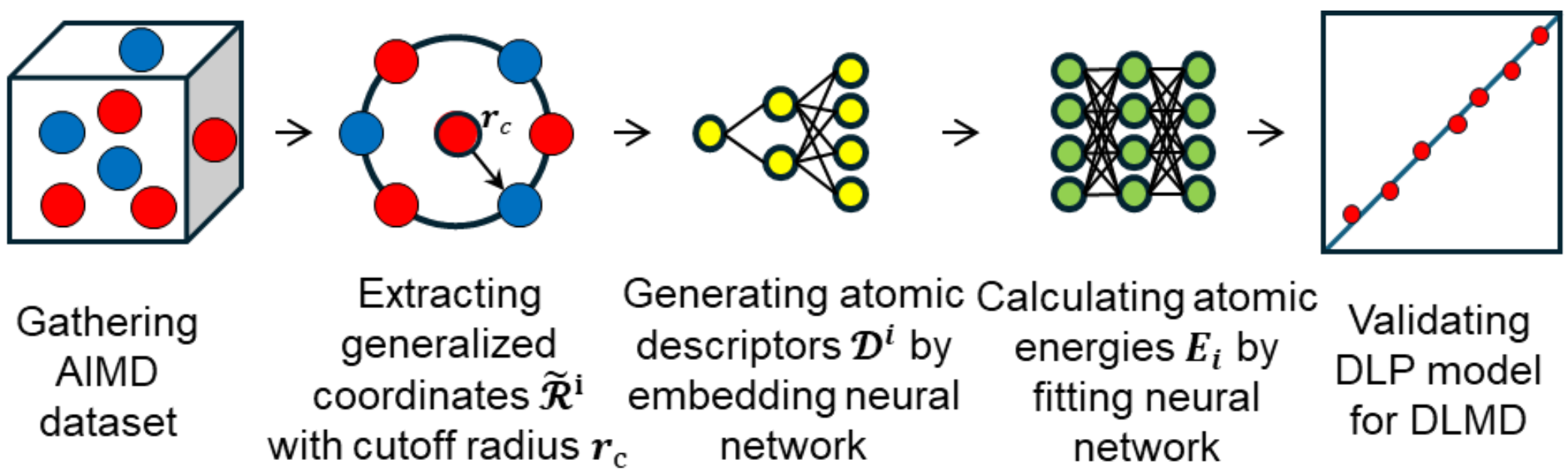

**Figure 1.** Development protocol of deep learning potential model for $Li_2ZrCl_6$.

More specifically, at the first step DFT calculations were performed using VASP, "Vienna Ab initio Simulation Package", version 5.4.4 to optimize initial structures [47]. The PAW, "Projector Augmented Wave" method [48] and the PBE, "Perdew-Burke-Ernzerhof" exchange-correlation functional [49] were employed. In order to obtain initial structures consistent with the actual structure in crystallographic parameters, the long-range van der Waals (vdW) interactions were accounted using Grimme's third-generation dispersion correction (DFT-D3) method with Becke-Johnson damping [50]. The electronic wave functions were expanded in a plane-wave basis set with an energy cutoff of 600 eV to achieve a balance between the convergence precision and computational cost. The valence electron configuration was set to $2s^22p^0$ for lithium, Li, $5s^24d^25p^0$ – zirconium,

Zr and $3s^23p^5$ – chlorine, Cl atoms. Although Zr is a transition metal, the Hubbard on-site Coulomb repulsion correction (U-parameter) was omitted because $Zr^{4+}$ has an empty 4*d* shell and, hence, no localized *d* electron subject to self-interaction errors [51][52].

Nevertheless, considering the effects of vdW interactions is crucial. By comparing calculations without vdW corrections and those employing different dispersion correction methods [50,53], we observed that the optimized atomic coordinates typically exhibit a higher root mean square deviation when vdW interactions are neglected. This difference is particularly pronounced in *β*-LZC. In addition, in terms of lattice parameters, the results obtained by the DFT-D3 method have the smallest deviation from the experimental findings of lattice parameters (**Table S1**). In particular, the change in the unit cell angle is negligible, which shows that the DFT-D3 method can keep the crystal geometry as consistent with reported data. Accordingly, the DFT-D3 method was selected as the optimal computational framework for subsequent simulations of trigonal *α*-LZC and monoclinic *β*-LZC structures.

The initial structures were based on experimental data obtained using Rietveld refinement results from room-temperature neutron powder diffraction [14,19]. A 1 × 1 × 1 unit cell with 6 Li, 3 Zr, and 18 Cl atoms was used to simulate the trigonal *α*-LZC structure. For the most stable ordered *α*-LZC, the optimized lattice parameters were determined to be $a$ = 10.827 Å, $b$ = 10.827 Å, $c$ = 5.818 Å, $\alpha$ = 89.970°, $\beta$ = 89.970°, and $\gamma$ = 120.387°. For the most stable disordered *α*-LZC, the optimized lattice parameters were determined to be $a$ = 10.875 Å, $b$ = 10.875 Å, $c$ = 5.863 Å, $\alpha$ = 90.000°, $\beta$ = 90.000°, and $\gamma$ = 120.000°. A 1 × 1 × 1 unit cell containing 4 Li, 2 Zr, and 12 Cl atoms was used to model the monoclinic *β*-LZC structure. For the most stable *β*-LZC, the optimized lattice parameters were determined to be $a$ = 6.288 Å, $b$ = 10.898 Å, $c$ = 6.280 Å, $\alpha$ = 90.000°, $\beta$ = 109.966°, and $\gamma$ = 90.000°. Through *k*-points density testing (**Figure S1**), 2 × 2 × 4 and 4 × 2 × 4 *k*-point grids were selected for Brillouin zone integration of *α*-LZC and *β*-LZC, respectively, to achieve the best balance between accuracy and computational efficiency. Both *α*- and *β*-LZC structures were fully optimized for atomic positions, cell shape, and cell volume until the force on each atom was reduced to less than 0.001 eV $Å^{-1}$. The side view and top view of the optimized structure are shown in **Figure 2**. For all LZCs, Zr atoms occupy octahedral sites coordinated by six Cl atoms, forming zirconium hexachloride $ZrCl_6^{-}$ octahedra with Cl atoms acting as ligands at the vertices of the octahedra. The $ZrCl_6^{2-}$ octahedra of ordered *α*-LZC are distributed in different *xy*-planes, representing a compact Zr layered structure. The $ZrCl_6^{2-}$ octahedra of disordered *α*-LZC are all located in a specific *xy*-plane, which means a single-layer crystallographic layer of Zr.

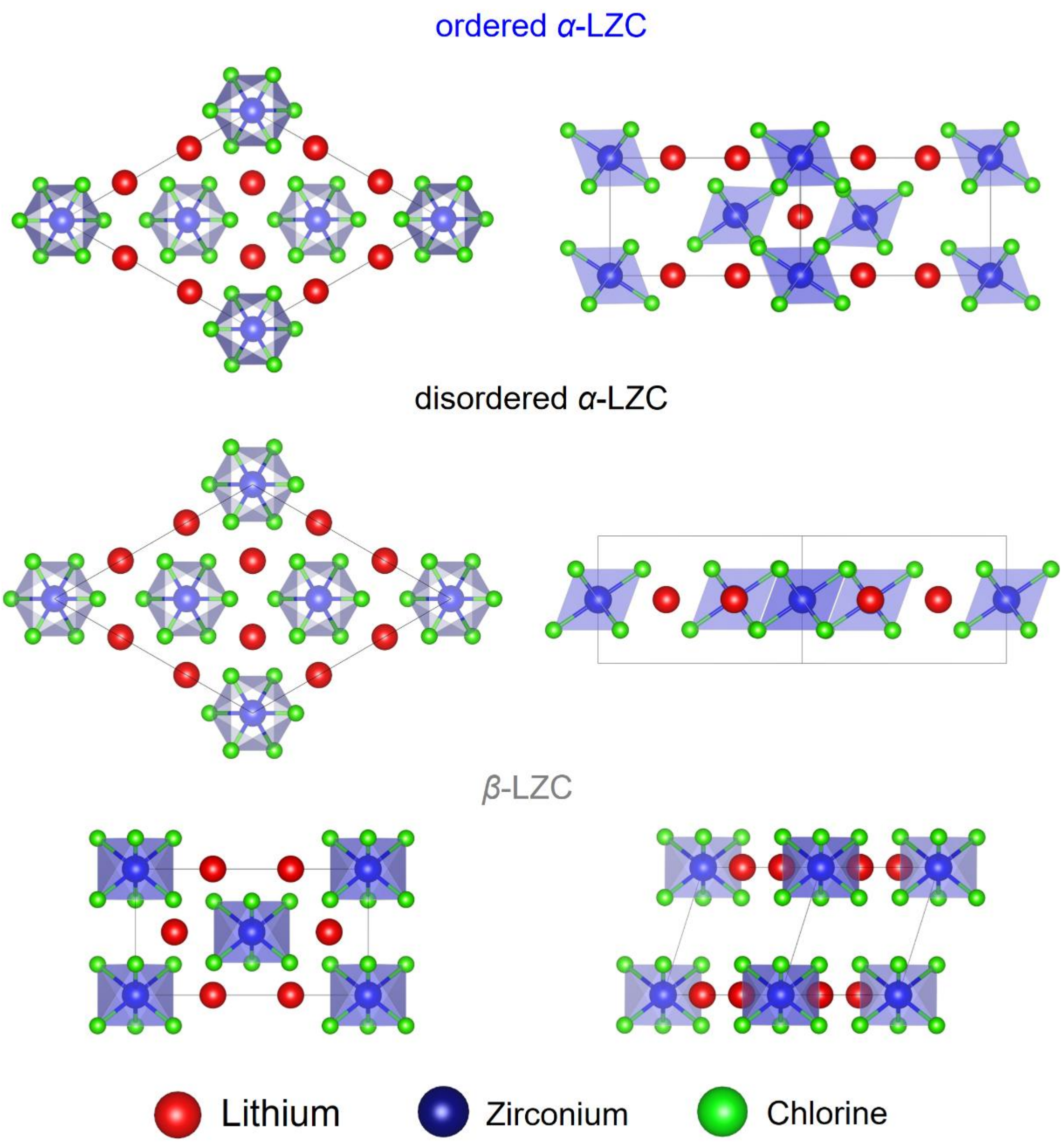

**Figure 2.** Optimized geometries to $Li_2ZrCl_6$, LZC, phases for ordered *α*-LZC (*top row*), disordered *α*-LZC (*middle row*) and *β*-LZC (*bottom row*) obtained by means of Density Functional Theory calculations, shown in top (*left column*) and side (*right column*) views. The solid black lines represent the 1 × 1 × 1 unit cell and light blue polyhedral represents zirconium hexachloride, $ZrCl_6^{2-}$, octahedra that forms a structural backbone of the LZC lattice (see Results and Discussion section).

Next, AIMD simulations were performed to generate the datasets for training DLP model. The simulations were performed at 300, 500, 700, 850, 1000, 1200, and 1400 K for the most stable optimized *α*- and *β*-LZC 1 × 1 × 1 unit cell structures for 5 ps under the canonical ensemble with a time step of 1 fs. By applying these temporal parameters, each temperature condition can yield up to 5,000 frames, thereby generating a robust set of physically consistent configurations suitable for training the DLP model with high efficiency. At the same time, this approach made it possible to develop the DLP model for different temperature ranges and account for the temperature fluctuations. In addition, to enhance the dataset diversity, AIMD simulations were performed for 0.1 ps at each temperature using rescaled coordinate structures with scaling factors of 0.75, 0.85, 0.95, 1.05, 1.15, and 1.25. Rescaling the coordinate structures enables a wider range of extreme configurations, including much larger or smaller interatomic distances, than those accessible in the original cell at a given temperature. Consequently, the trained DLP model is equipped with a more comprehensive potential energy surface (PES), thereby ensuring the stability of the simulation during DLMD processes when encountering transiently non-physical atomic distances. With regard to more detailed parameters used in AIMD, the same $k$-points grids and treatment of valence electrons as for DFT optimization were used. A kinetic energy cutoff of 450 eV was set to balance the computational resources with sufficient computational accuracy.

Based on the obtained AIMD trajectories, the DLP models were developed using a TensorFlow [54] deep neural network approach implemented in DeePMD-kit, version 2.2.9 [55]. Depending on the specific system, different cutoff radius $r_c$ of 6 Å (ordered $\alpha$-LZC) and 8 Å (disordered $\alpha$-LZC and $\beta$-LZC), as well as smooth cutoff parameter $r_{cs}$ of 0.7 Å were applied to design a two-body embedding smoothed version of deep potential (DeePot-SE) and end-to-end machine learning PES scan model. These parameters were chosen with a comprehensive consideration of the structural properties of LZCs and the need for computational efficiency to accurately characterize each system. Notably, DLP models built on this approach have ability to effectively represent the PES of a wide range of systems with the accuracy of *ab initio* calculations [55]. More details regarding to DLP training theory could be found in support information (**Section S1**).

The training process of the DLP model was designed to run 150,000 steps to ensure sufficient convergence of the complex PES based on training dataset. The training utilized the mean squared error of energy, forces, and the virial tensor as the loss function to simultaneously optimize the predictive accuracy for these three quantities [54,[55]. This choice of a multi-objective loss function reflected the critical physical correlation between energy and forces in LZCs, effectively capturing many-body interactions and dynamic behavior. The optimization process employed the Adam optimizer, starting from a learning rate of $1 \times 10^{-3}$ and gradually decaying to $3.51 \times 10^{-8}$, with the decay parameter set to 5,000. This learning rate scheduling strategy, which is the default for the Adam optimizer, facilitates the exploration of the global optimum through rapid parameter updates in the initial stages, followed by fine-tuning via a gradually decreasing learning rate, thereby enhancing the fitting accuracy to the complex PES in the later stages of training.

The training dataset comprised over 60,000 samples, while the testing dataset included over 10,000 samples, both extracted from various AIMD trajectories. Note that the two datasets are independent of each other to ensure that the validation process will not be contaminated by the training data. This approach ensured data diversity and broad coverage of the configuration space of the system. During the validation process, loss function parameters, including mean absolute error and root mean square error, were calculated (as displayed in **Table S2**) to quantify the predictive performance of the model based on testing dataset. The results demonstrated that the prediction accuracy for energy, forces, and the virial tensor reached approximately 99%, underscoring the efficiency of the selected parameters. Force comparison of the predicted data points with the reference data for Li, Zr and Cl atoms is shown in **Figures 3** (see **Figure S2** for energy and virial comparison). These findings together confirm the rationality of the training parameters and 1 × 1 × 1 unit cell structures choice in achieving a balance between model accuracy, stability, and computational efficiency.

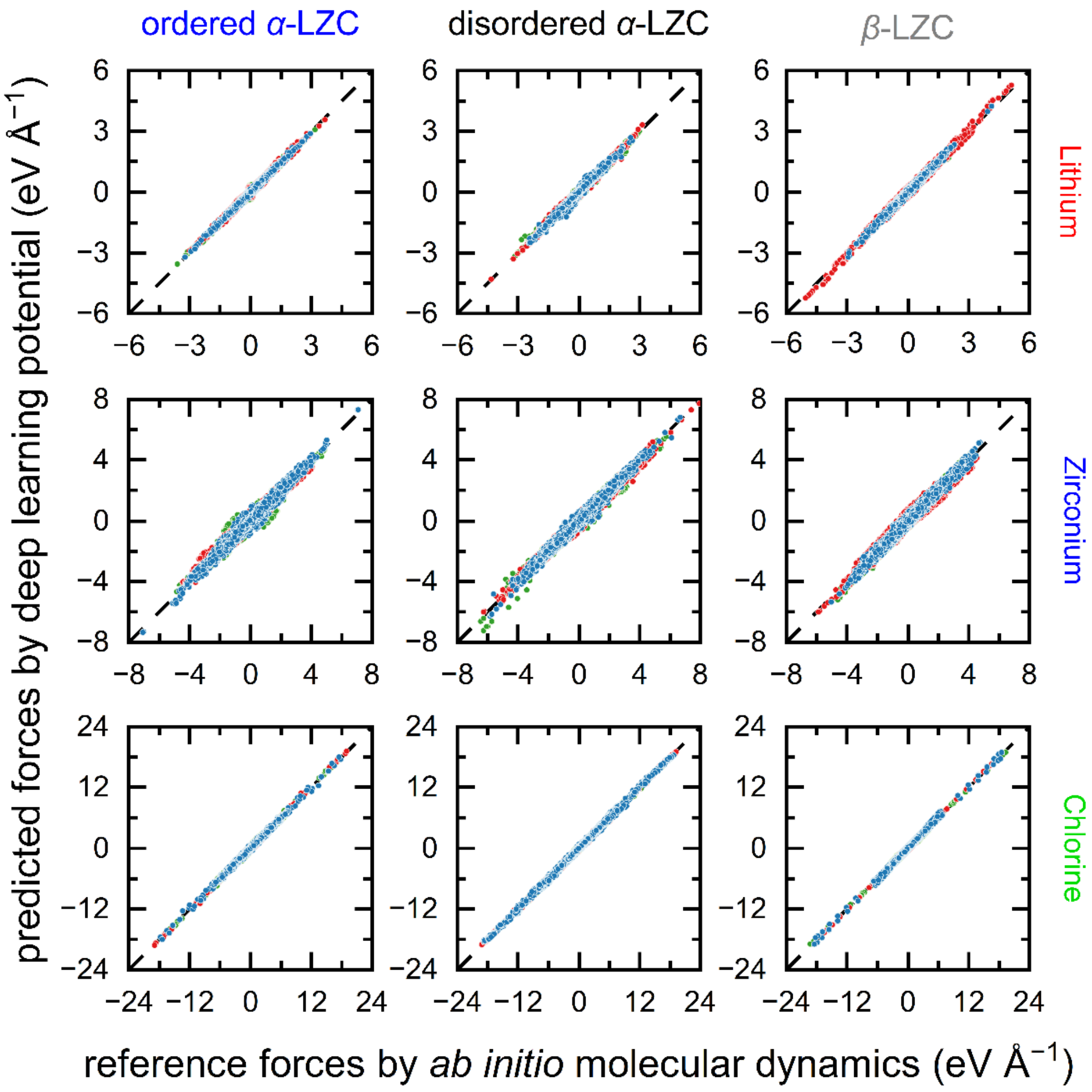

**Figure 3.** Validity of deep learning model based on testing dataset for $Li_2ZrCl_6$, LZC, for ordered $\alpha$-LZC (*left column*), disordered $\alpha$-LZC (*middle column*) and $\beta$-LZC (*right column*) phases, expressed by correlation between directional atomic forces predicted by deep learning potential and reference *ab initio* molecular dynamics simulation for lithium (*top row*), zirconium (*middle row*), chlorine (*bottom row*) atoms. *Green circles* represent forces in $x$-direction *red* – $y$-direction, *blue* – $z$-direction. In all cases, the estimated determination coefficient, $R^2$ = 0.99, indicates that 99% of the variance in *ab initio* forces is accurately captured by the deep learning model. *Black dashed line* stands for visual guidance and represent ideal correlation with $R^2$ = 1.

After obtaining well-trained DLP models, the DFT-optimized structures were replicated to construct the initial configurations for DLMD simulations. The simulation supercells were designed to contain 12,960 atoms for $\alpha$-LZC and 9,000 atoms for $\beta$-LZC with replication as 8 × 6 × 10 and 10 × 5 × 10, respectively. The selection of the supercell size and the number of atoms ensures adequate statistical sampling in terms of temporal and spatial resolution. Finally, DLMD simulations were performed in the same thermodynamic conditions as in AIMD stage using LAMMPS, "Large-scale Atomic/Molecular Massively Parallel Simulator" code, version 080223 [56], coupled with the DeepMD [55] plugin. For all the calculations the time step was set 1 fs. This time step ensures numerical stability while accurately capturing the rapid dynamic behavior of Li, Zr, and Cl at high temperatures. Temperature coupling was achieved using Nosé-Hoover [57],[58] thermostat at 750, 800, 850, 900, 950, 1000 K for $\alpha$-LZC and 900, 950, 1000, 1050, 1100, 1150 K for $\beta$-LZC. The damping parameter was set as 100 fs to balance the sampling of temperature fluctuations with the thermal response of the system, ensuring that the heat bath effectively regulated the system temperature without excessively perturbing the dynamical behavior. To gain a reasonable statistic, five independent DLMD simulations were performed for each system by varying the initial velocity seed at the beginning of the DLMD setup. Each DLMD simulation was conducted for at least 5 ns, providing a sufficiently long time window to capture the equilibrium thermodynamic properties and dynamic evolution. Energy variations during simulation are shown in **Figure S3.** For simplicity, only the 950 K trajectories are used as examples. It can be observed that all LZCs reach equilibrium quickly (less than 100 ps) and converge under a stable energy range (less than 0.01 eV atom$^{-1}$). These thermodynamically stable trajectories are used in subsequent analysis.

To thoroughly investigate the transport mechanism in LZCs we conducted a systematic analysis of the DLMD trajectories using a stepwise approach by employing the mean-squared displacement to validate experimental data and quantify ion dynamics, van Hove correlation function and residence time to characterize transport mechanism, spatial and radial distribution function to probe structural environment and migration pathways, as illustrated in **Figure 4**.

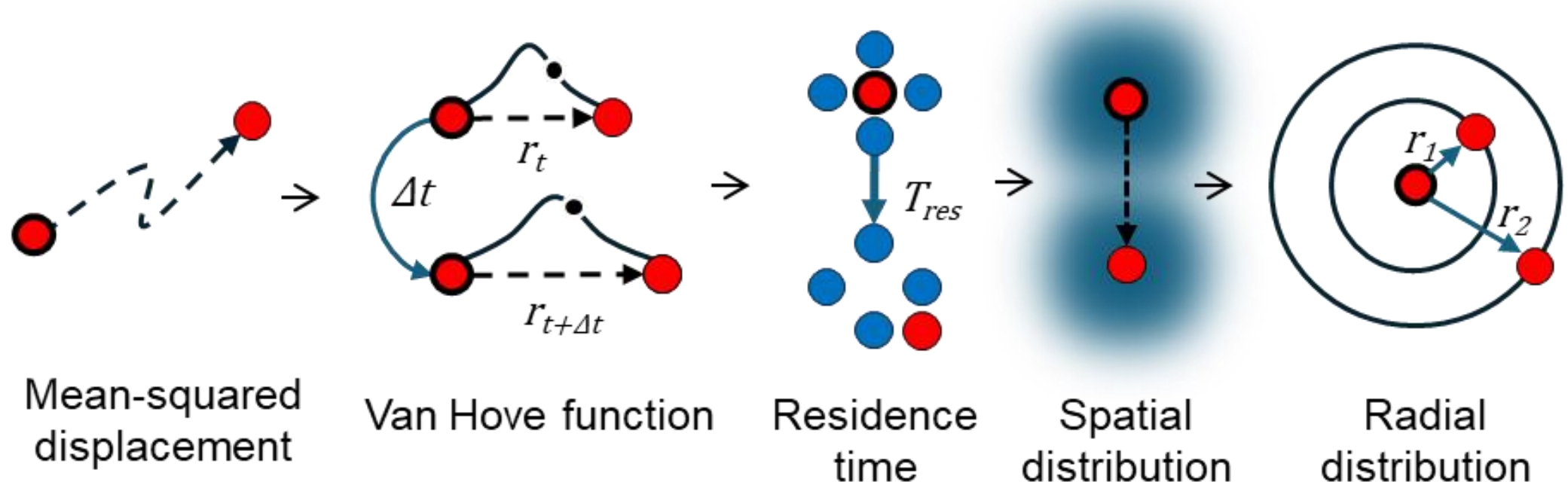

**Figure 4.** Analytical protocol to reveal the ion transport mechanism in $Li_2ZrCl_6$.

Particularly, the diffusion of lithium ions was determined using mean-squared displacement, $\boldsymbol{r}^2(t)$, where the positions of lithium $i$ and $j$ relative to the center-of-mass of the system, $\boldsymbol{r}_i$ and $\boldsymbol{r}_j$ respectively, were tracked as a function of time $t$ [59,60]:

$$\boldsymbol{r}^2(t) = \sum_i \sum_j \langle [\boldsymbol{r}_i(t) - \boldsymbol{r}_i(0)][\boldsymbol{r}_j(t) - \boldsymbol{r}_j(0)] \rangle \quad (1)$$

here brackets $\langle \cdots \rangle$ indicate an average over the ensemble of particles and time. These collective displacements consist two contributions: when $i = j$, the auto-correlation function of the $\alpha$ particle flux is considered, whereas for $i \neq j$, the cross-correlation between two different particle fluxes is calculated. These two contributions can be expressed as the individual ion displacement (or self-displacement), $\boldsymbol{r}_i^2(t)$, and the correlated (distinct-displacement) terms, $\boldsymbol{r}_{ij}^2(t)$, respectively:

$$\boldsymbol{r}_i^2(t) = \sum_{\alpha} \langle [\boldsymbol{r}_i(t) - \boldsymbol{r}_i(0)]^2 \rangle \tag{2}$$

$$\boldsymbol{r}_{ij}^2(t) = \boldsymbol{r}^2(t) - \boldsymbol{r}_i^2(t) = \sum_{i} \sum_{j \neq i} \langle [\boldsymbol{r}_i(t) - \boldsymbol{r}_i(0)][\boldsymbol{r}_j(t) - \boldsymbol{r}_j(0)] \rangle \tag{3}$$

In these equations, when $\boldsymbol{r}_{ij}^2(t) > 0$, the motion between particles is correlated and has a positive impact on particle transport. Due to the typically large errors between different terms, an appropriate data averaging is essential to obtain reliable expected values. Since increasing the system size and extending simulation time do not effectively yield suitable and smooth displacements, the proper way is to average the data from multiple different trajectories to achieve reasonable values [61].

The diffusion coefficient, $D$, was then calculated using the Einstein formalism:

$$D = \frac{\boldsymbol{r}^2(t)}{2dt} \tag{4}$$

where $d$ is a dimensionless factor equal to 3, corresponding to three-dimensional displacement. The diffusion was estimated based on Fickian formalism, according to which $\boldsymbol{r}^2(t) \infty t^{\beta}$ where $\beta$ coefficient was estimated using the equation:

$$\beta = \frac{dlog(\boldsymbol{r}^2(t))}{dlog(t)} = 1 \tag{5}$$

and $\beta = 1$ represents the slope of 45° in the time window where diffusion regime occurs.

To quantify the degree of correlated motion between the particles during their diffusion, the Haven's ratio, $H_R$ **Error! Reference source not found.**,63], was estimated as the ratio between the diffusion of individual particles, $D_i$, and the collective diffusion coefficient, $D$:

$$H_R = \frac{D_i}{D} \tag{6}$$

In scenarios where particle dynamics are uncorrelated, $D_i \equiv D$, and $H_R$ reaches a value of 1. Conversely, if $H_R$ < 1, this indicates the presence of the correlation effects in the system.

To validate DLP model, ionic conductivity was determined based on the results obtained from **Equation (7)** using the Onsager approximation [64]:

$$\sigma = \frac{ce^2Z^2}{k_BT}D = \frac{ce^2Z^2}{H_Rk_BT}D_i \quad (7)$$

here $c$ represents the number of lithium ions per unit volume, $e$ denotes the elementary charge, $Z$ corresponds to the valence state of particle equal to one, and $k_B$ stands for the Boltzmann constant. Due to significantly reduced dynamics of solid-state electrolyte at room temperature, ionic conductivity obtained at the high temperatures was used to estimate the ionic conductivity under experimental thermodynamic conditions through the Arrhenius relationship [22,65]. It is worth noting that the activation energy is generally considered to be a constant value. Therefore, the activation energy of the system was additionally determined by analyzing the ionic conductivity data at different temperatures:

$$ln\,(\sigma) = ln\,(\sigma_0)\; - \;\frac{E_a}{k_B}\frac{1}{T} \quad (8)$$

By performing a linear regression of the ionic conductivity as a function of the reciprocal temperature $T$, the activation energy $E_a$ of the corresponding system was obtained by multiplying the negative of the slope by the Boltzmann constant $k_B$.

In addition, van Hove correlation function was used to describe the correlation between two species $i$ and $j$ in time and space:

$$G(\boldsymbol{r},t) = \frac{1}{N}\langle\sum_{i=1}^{N}\sum_{j=1}^{N}\delta\left(r - \boldsymbol{r}_i(t) + \boldsymbol{r}_j(0)\right)\rangle \quad (9)$$

where $r$ is radial distance in space, $t$ is the time interval from the initial moment to the observation moment; $N$ is total number of particles. Based on **Equation (9)**, the individual particle $i$ displacements were analyzed to reveal the self ion transport mechanism:

$$G_s(\boldsymbol{r},t) = \frac{1}{N}\langle\sum_{i=1}^{N}\delta\left(r - \boldsymbol{r}_i(t) + \boldsymbol{r}_i(0)\right)\rangle \quad (10)$$

To gain deeper insight into the particle transport, the coordination dynamics analysis was applied to describe the atomic residence time under the spatial confinement:

$$\beta_{ij}(t) := \begin{cases} 1 & if\ criteria\ fulfilled \\ 0 & otherwise \end{cases} \quad (11)$$

$\beta_{ij}(t) = 1$ when atom $i$ is in proximity to atom $j$. Using this method, the autocorrelation function $c(\tau)$ which is used to express correlation of the dwell state after time $\tau$ and residence time $T$, can be computed:

$$c(\tau) = \frac{1}{N^2}\sum_{i=1}^{N}\sum_{j=1}^{N}\int_{0}^{\infty} \beta_{ij}(t)\,\beta_{ij}(t+\tau)\,dt \tag{12}$$

$$T = 2\int_{0}^{\infty} c(\tau)d\tau \tag{13}$$

here $N$ and $\tau$ are total atoms to be considered and time within the proximal distance, respectively.

Based on the **Equation (4)** and **(13)**, the hopping dynamics was analyzed by the Chudley-Elliott model [66] to quantify the average ion jump-length, *l*:

$$l = \sqrt{6D_i\tau} \tag{14}$$

To confirm the accuracy of the structural features of LZC, X-ray diffraction was applied to compare simulation structures with experimental results:

$$n\lambda = 2d_{hkl}\sin\theta \tag{15}$$

$$F_{hkl} = \sum_{j} f_j e^{2\pi i(hx_j+ky_j+lz_j)} \tag{16}$$

$$I_{hkl} \propto |F_{hkl}|^2 \cdot LP \tag{17}$$

where $n$, $\lambda$, and $d_{hkl}$ represent the diffraction order, wavelength of X-rays and interplanar spacing, respectively. In order to match the experimental conditions, Cu $K_{\alpha 1}$ was used as the X-ray radiation source. Therefore, $\lambda = 1.5406$ Å. $d_{hkl}$ is determined by reciprocal lattice of crystal with miller index $h$, $k$ and $l$. $\theta$ is Bragg angle which means half angle of incidence and diffraction. Diffraction intensity $I_{hkl}$ is been calculated by structure factor $F_{hkl}$ and Lorentz-polarization factor $LP$. Here $f_j$ and $x_j$, $y_j$, $z_j$ mean atomic scattering factor and fractional coordinates the atom $j$ in the unit cell.

To describe the structural features of the studied systems, the radial distribution function analysis was employed:

$$g(r) = \frac{1}{N}\sum_{i=1}^{N}\sum_{j\neq i}\frac{1}{4\pi {r_{ij}}^2\Delta r}\delta(r - r_{ij}) \tag{18}$$

where $N$, $r_{ij}$, and $\Delta r$ represent the total number of atoms within a radius $r$, the distance between atoms $i$ and $j$, and the bin width, respectively. Term $\Delta r$ is set to 0.032 Å, which is significantly smaller than the typical interatomic distances in the system (1-2 Å). This value was chosen to strike an optimal balance between achieving high spatial resolution, sufficient to resolve the fine structural details of the nearest coordination shell and reduce the statistical errors – noise – in the resulting distribution.

To direct demonstrate the energetic pattern, the potential of mean force was calculated to elucidate the Gibbs free energy landscape governing atom migration in LZC phases:

$$W(r) = -k_B T \ln g(r) \tag{19}$$

$$\Delta G = W(r_{max}) - W(r_{min}) \tag{20}$$

where $k_B$ and $T$ are the Boltzmann constant and temperature. Gibbs free energy barrier $\Delta G$ is the potential of mean force gap between local maximum $W(r_{max})$ – transition state – and local minimum $W(r_{min})$ – corresponding to the steady state. In this way, we can observe the free energy barriers of particle diffusion at different temperatures from a more comprehensive thermodynamic perspective. For more concise description, we will use energy barrier instead of Gibbs free energy barrier in the following article.

Finally, as a further proof of the energy barrier, the density of states and crystal-orbital Hamilton populations were calculated to illustrate the effect of the electronic structure differences in different LZCs phases on the energy barrier:

$$D(E) = \frac{1}{V} \sum_{i=1}^{N} \delta(E - E(k_i)) \tag{21}$$

where $V$ and $N$ are the volume and number of *k*-points in the discrete grid sampling Brillouin zone. $\delta(E - E(k_i))$ is used to count the states where the energy of the $i$-th *k*-point ($E(k_i)$) is equal to $E$. Partial density of states can be obtained by decomposing $D(E)$ into contributions from specific atoms, orbitals (such as *s*, *p*, *d*), or subspaces.

$$COHP_{ij}(E) = H_{ij}(E) \cdot D_{ij}(E) \tag{22}$$

$$ICOHP_{ij}(E) = \int_{-\infty}^{E_f} COHP_{ij}(E) dE \tag{23}$$

where $COHP_{ij}(E)$ and $ICOHP_{ij}(E)$ are crystal-orbital Hamilton populations and their integrals for atomic pair $i, j$. $H_{ij}(E)$ and $D_{ij}(E)$ are Hamiltonian matrix elements and density of states contributions.

The results for **Equation (1)-(8)** and **Equation (19)-(20)** were obtained using in-home Python code, **Equations (9)-(14)** and **Equations (18)** were analyzed under the TRAVIS, "TRajectory Analyzer and VISualizer", code, version 062922 [67]. **Equations (15)-(17)** were calculated by Pymatgen, "Python Materials

Genomics", library [68]. **Equation (21)** was determined by VASPKIT [69]. **Equation (22)-(23)** were evaluated from LOBSTER [70].

## Results and Discussion

Before proceeding with the analysis of molecular dynamics trajectories, we first aimed to validate the developed deep learning potential (DLP) model of $Li_2ZrCl_6$ (LZC) extending the assessment beyond the convergence of predicted and reference *ab initio* forces. For this purpose, we employed radial distribution functions analysis (**Equation (18)**) to compare the spatial arrangement of the atoms using the spatial and temporal scales matching *ab initio* molecular dynamics simulation at room temperature, represented by 300 K and high temperature, represented by 950 K (**Figure S4**). To maintain proper variable control, all conditions except the force field, including the initial structure, time step, and trajectory length (1 ps), were kept constant. The obtained results demonstrate that the peak intensities, positions and shapes are in good agreement with each other, suggesting the high degree of consistency between two methods. Furthermore, this indicates the preservation of the crystal structure across the wide temperature range and capability of DLP model to accurately predicting the static properties of the system under different thermodynamic conditions. At the second step, the DLP model was further validated by its ability to reproduce the dynamic properties by evaluating the ionic conductivities and corresponding activation energies for the different LZC phases and comparing our results with available experimental data. To achieve statistically meaningful and reliable mean-squared displacement evolutions, each trajectory was divided into five equal segments, yielding 20 trajectory segments per system, which were further averaged to reduce the statistical uncertainty and enhance the accuracy of slope determination within the diffusive regime (**Figure 5a**). Thus, the estimated time window for diffusive regime (**Equation (5)**), used for calculating ionic conductivities (**Equation (7)**) and further activation energies (**Equation (8)**), was identified between approximately 300 and 700 ps as illustrated in **Figure S5**. For simplicity, only 950K was taken as an example, where the displacement exhibited clear linear behavior. Such a meticulous approach avoids vibrational or sub-diffusive motion contributions and, therefore, avoids the overestimation or underestimation of the target quantities.

Upon preliminary visual analysis of the resulting displacements curves reveals a progressive increase with temperature, assuming enhancing lithium-ion mobility, which is proportional to the ionic conductivity, at higher temperatures. This temperature-dependent behavior aligns with expected Arrhenius-type dynamics observed in solid-state electrolytes [16,22,31,32]. It is important to note that from a computational perspective ion dynamic at room temperature are considerably slow. Therefore, to reach the diffusive regime and, hence, statistically reliable values of ionic conductivity, significantly long-timescale simulations are required. For example, preliminary simulations show that ordered $\alpha$-LZC at 600 K and $\beta$-LZC at 800 K unable to reach the diffusive regime within 5 ns, while higher simulation temperatures (750 K for ordered $\alpha$-LZC and 900 K for $\beta$-LZC) can provide sufficient ion dynamics (**Figure S6**). At the same time, we also consider 750 K as the minimum simulation temperature to disordered $\alpha$-LZC for reaching the diffusive region based on the similarity of ionic conductivity with ordered $\alpha$-LZC. To overcome this temperature limitation, we applied the Arrhenius relationship to extrapolate the ionic conductivity at 300 K and to obtain the corresponding activation energy for each of the investigated LZC phases. The temperature dependence of ionic conductivity obtained at higher temperatures (**Figure 5b**) served as a reliable basis for this extrapolation, enabling accurate estimation of room-temperature ionic transport properties comparable with previously reported experimental measurements for $Li_2ZrCl_6$ [14-21]. Thus, the estimated ionic conductivities and activation energies were found to be in good agreement with experimental data (**Table 1**) with the discrepancies not exceeding *ca.* 22% and even far exceed the previously obtained results using the *ab initio* molecular dynamics (more than 80%), confirming the validity and predictive ability of our computational approach and DLP model (see **Table S3** for further details on numerical data for ionic conductivity). Meanwhile, it is worth noting that since ordered and disordered $\alpha$-LZC coexist in the experiments [17], the ionic conductivity and activation energy for ordered and disordered $\alpha$-LZC are combined and expressed as the same value.

**Table 1.** Lithium-ion transport characteristics in $Li_2ZrCl_6$, LZC, expressed by individual and collective diffusion coefficient, $D_i$ and $D$ (cm$^2$ s$^{-1}$), Haven's ratio, $H_R$, ionic conductivity, $\sigma$ (S cm$^{-1}$), and activation energy, $E_a$ (eV), estimated by extrapolation to 300 K for ordered and disordered $\alpha$-LZC and $\beta$-LZC by means of deep learning-accelerated molecular dynamics in comparison with experimental data [14-21] and previous simulation results based on *ab initio* molecular dynamics [16,22,31,32].

| | $D_i$ | $D$ | $H_R$ | $\sigma$ | | | $E_a$ | | |
|---|---|---|---|---|---|---|---|---|---|
| | this work | | | this work | experiment | other simulation | this work | experiment | other simulation |
| ordered $\alpha$-LZC | $4.18\times10^{-9}$ | $1.04\times10^{-8}$ | 0.40 | $6.57\times10^{-4}$ | $0.98\text{-}8.08\times10^{-4}$ | $3.21\text{-}5.56\times10^{-3}$ | 0.33 | 0.32-0.42 | 0.20-0.29 |
| disordered $\alpha$-LZC | $2.82\times10^{-9}$ | $2.09\times10^{-8}$ | 0.13 | $1.32\times10^{-3}$ | | $0.54\text{-}12.20\times10^{-4}$ | 0.30 | | 0.30-0.38 |
| $\beta$-LZC | $1.21\times10^{-10}$ | $1.25\times10^{-10}$ | 0.97 | $7.66\times10^{-6}$ | $5.70\text{-}7.10\times10^{-6}$ | $1.00\times10^{-6}$ | 0.45 | 0.36-0.50 | 0.53 |

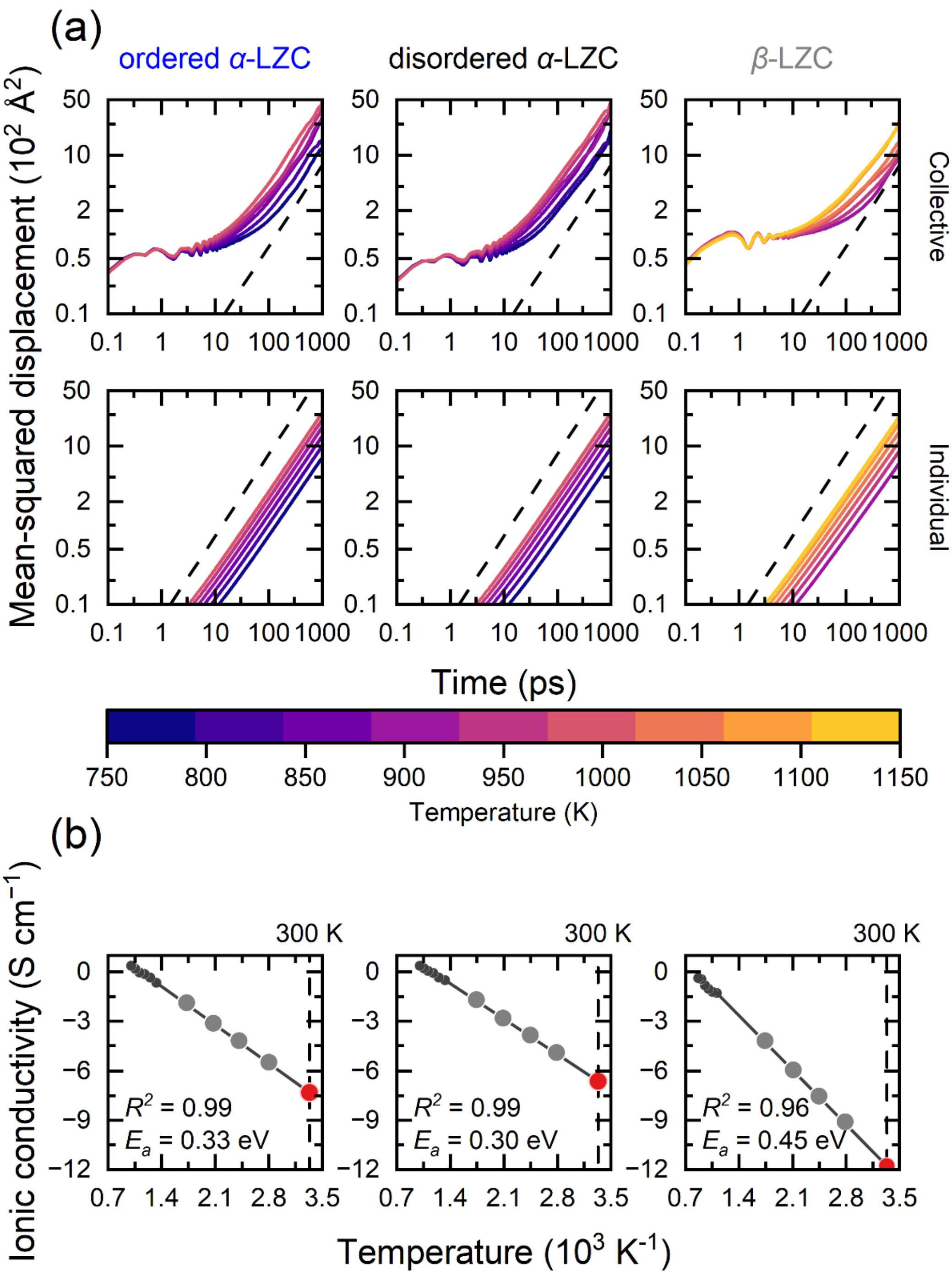

**Figure 5.** Lithium-ion transport characteristics in $Li_2ZrCl_6$, LZC, for ordered *α*-LZC (*left column*), disordered *α*-LZC (*middle column*) and *β*-LZC (*right column*) phases, expressed by **(a)** collective (*top row*) and individual (*bottom row*) mean-squared displacement of lithium ions obtained by the averaging over 20 runs in temperature range between 750 and 1000 K for *α*-LZC and between 900 and 1150 K for *β*-LZC. Both axes are plotted in logarithmic scales. *Black dashed line* drawn at a 45° stands for visual guidance and represent the time window at which diffusion regime occurs (see **Figure S5, S7** for additional details); **(b)** Arrhenius plots of ionic conductivity and corresponding activation energy, $E_a$. The ionic conductivity is plotted on natural logarithmic scale as a function of inverse temperature. The *grey circles* stand for visual guidance of ionic conductivity values at different temperatures and *red circles* indicate the extrapolated value of ionic conductivity at 300 K. The estimated determination coefficient, $R^2$ = 0.99 and 0.96, indicates that 99% and 96% of the variance in the temperature dependent ionic conductivity is well captured by the deep learning model.

Building on the mean-squared displacement and ionic conductivity analysis, a comparative analysis of displacement magnitudes across the three structures can also highlight a significant difference in lithium migration. Indeed, the disordered $\alpha$-LZC consistently shows the highest displacement values compared to $\beta$-LZC, suggesting the origin of the difference in ionic conductivity. Further comparison between collective (**Equation (1)**), comprising self- and distinct-correlation terms, and individual ion displacements (**Equation (2)**) may provide additional and deeper insights into the presence and magnitude of ionic correlation effects. The larger discrepancies between collective and individual displacement observed in $\alpha$-LZCs may be a sign of stronger correlation effects contributing substantially to ionic conduction, whereas the similarity between the displacements for $\beta$-LZC may indicate that ion transport is predominantly governed by uncorrelated rather than by correlated motion. To quantify these observations, we further estimated the Haven's ratio (**Equation (6)**), which measures the degree of correlated ionic motion. The computed Haven's ratio values for the LZC phases were found to be consistently below unity, particularly for $\alpha$-LZC structures, confirming the presence of significant interionic correlation effects. This result indicates the necessity of considering correlated contributions, as relying solely on individual ionic motion would underestimate the ionic conductivity and activation energies. In contrast, for $\beta$-LZC the degree of the correlated motion approached unity, indicating that correlation effects are minimal, and individual ionic dynamics dominates in overall transport behavior. This interpretation aligns with the more pronounced linear curves observed for $\alpha$-LZCs correlated displacements (**Figure S7**), whereas the noisier, less-defined profile for $\beta$-LZC curves reflects weaker correlation effects. The correlated ionic motion in $\alpha$-LZCs can be attributed to mutual lithium-ion repulsion at close spatial proximities, which enhancing overall ionic mobility and conductivity. These findings are consistent with simulation results reported in previous studies [22], which demonstrate that correlated effects in lithium-ion diffusion significantly reduce migration barriers. This corroborates the critical importance of incorporating correlated effects into the analysis of ionic conductivity and activation energies, further validating the accuracy of our study in elucidating the underlying physical mechanisms.

Our analysis further reveals that $\beta$-LZC exhibits the lowest ionic conductivity and highest activation energy among the investigated phases, which aligns closely with available experimental findings. Conversely, the disordered $\alpha$-LZC structure demonstrates the highest ionic conductivity and lowest activation energy, indicating superior lithium-ion transport capability. Beyond ionic correlation effects, we also hypothesized that the observed differences in ionic conductivity and activation energies primarily originate from different lithium-ion diffusion modes. Particularly, lithium-ion diffusion in $\alpha$-LZC phases exhibits anisotropy, with interlayer diffusion rates significantly surpassing intralayer diffusion. In contrast, the $\beta$-LZC phase displays isotropic diffusion, characterized by a greater interlayer diffusion rates than the intralayer diffusion rates. Such spatial dependency may directly influence ion transport and contribute significantly to the variation in dynamic properties across these electrolyte systems. This hypothesis is supported by prior experimental conjectures and corroborated by bond valence site energy analyses [14,20]. Previous experimental work identified a favorable 3D percolating network in $\alpha$-LZC with a low effective migration barrier of 0.803 eV, driven by interlayer Li displacements along the [001] direction, and a higher effective barrier of up to 0.809 eV in *β*-LZC, with Li displacements primarily within the *xy*-plane.

To support our hypothesis about the lithium transport behavior in LZCs, we decomposed the collective and individual ion displacements into directional components along the *x*-, *y*-, and *z*-axes estimated under identical thermal conditions at 950 K. The trajectory temperature is chosen to be 950 K because it falls within the overlapping temperature ranges tested for *α*-LZC (750, 800, 850, 900, 950, 1000 K) and *β*-LZC (900, 950, 1000, 1050, 1100, 1150 K), facilitating direct comparison of lithium-ion transport behavior between the two phases. Compared to room-temperature simulations, high-temperature simulations significantly accelerate lithium-ion diffusion, enabling statistical equilibrium within the simulation timescale and clearly elucidating diffusion pathways and correlated dynamics. Crucially, the high determination coefficient obtained from the Arrhenius relationship confirms that there is no phase transitions affecting diffusion behavior at 950 K. The observed diffusion trends are consistent with those expected at room temperature, therefore, 950 K was chosen as the representative temperature for trajectory analysis of LZCs. Upon precise examination of the resulting displacements in different spatial directions (**Figure 6**), we observed a significant anisotropy in lithium transport within both *α*-LZC structures. In particular, the displacements along *z*-direction exhibit higher values compared to the sum of *x* and *y* for both collective and individual components. This observation suggests that lithium diffusion in *α*-LZCs is predominantly governed by interlayer (along *z* direction) transport rather than intralayer (*xy*-plane, sum of *x* and *y* directions) diffusion. In contrast, *β*-LZC phase demonstrates a more uniform distribution of displacements across all three directions, hence indicating isotropic lithium transport behavior and higher intralayer diffusion ratio. Further examination of directional components provides additional insight into the impact of correlated contribution into the collective transport. As illustrated in **Figure S8**, the *α*-LZC structures not only exhibit a strong interlayer diffusion tendency in the individual motions but also show even more pronounced anisotropic tendency in correlated dynamics. Notably, almost all of the correlated diffusion contributions in *α*-LZC arise along the *z*-direction, further highlighting the tendency of strong interlayer correlated motion. For *β*-LZC, the relatively weak lithium correlation effects results in collective displacements closely resembling the individual ones with minimal directional deviation.

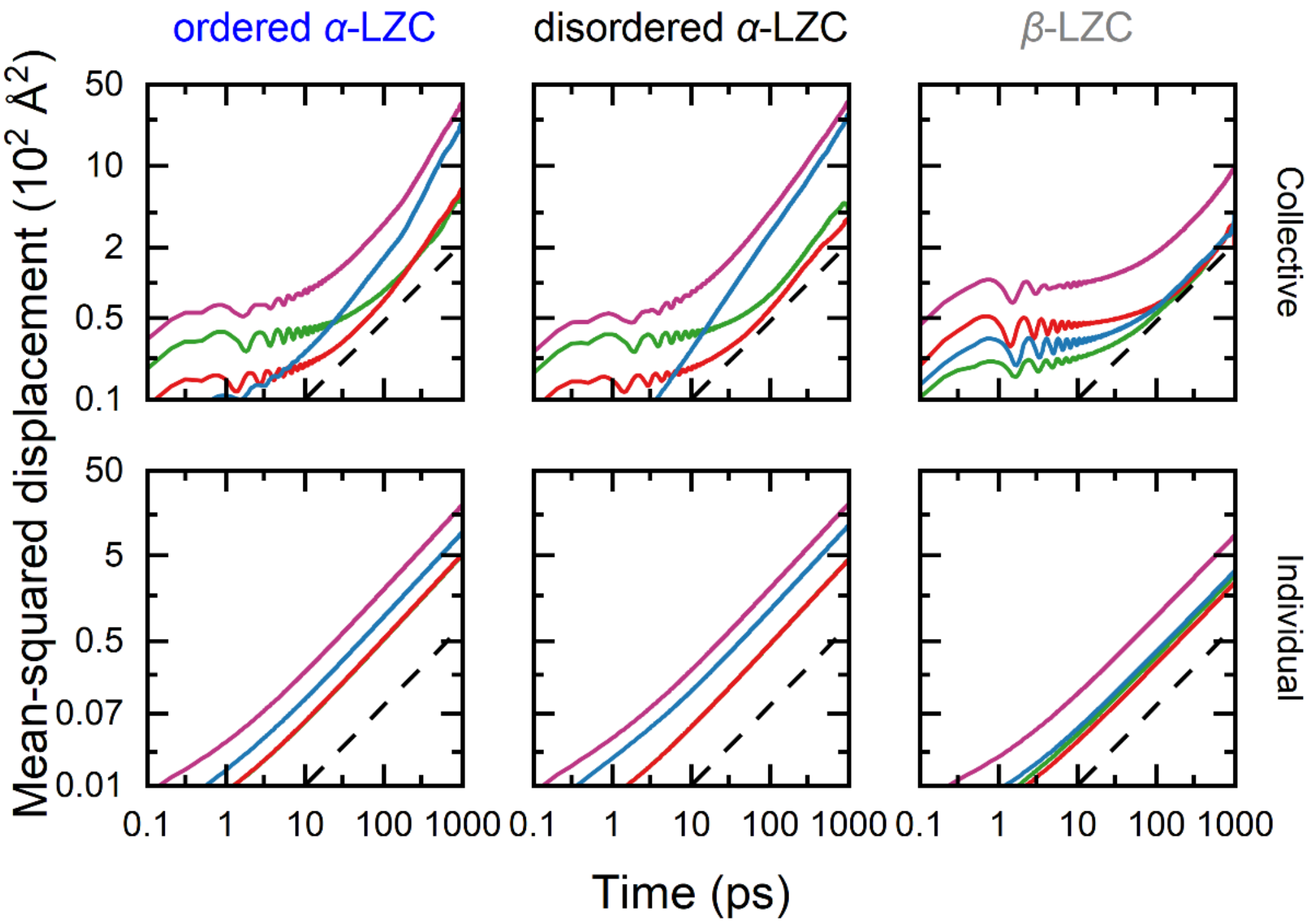

**Figure 6.** Lithium-ion transport characteristics in $Li_2ZrCl_6$, LZC, for ordered *α*-LZC (*left column*), disordered *α*-LZC (*middle column*) and *β*-LZC (*right column*) phases, expressed by collective (*top row*) and individual (*bottom row*) mean-squared displacement of lithium ions with cumulative (*pink line*) and directional, *xyz*, terms calculated at 950 K by means deep learning-accelerated molecular dynamics simulation. *Green lines* represent displacement in *x*-direction *red* – *y*-direction, *blue* – *z*-direction. Both axes are plotted in logarithmic scales. *Black dashed line* drawn at 45° stands for visual guidance and represent the time window at which diffusion regime occurs (see **Figure S8** for corresponding linear representations).

Quantifying these observations with ionic conductivity analysis may further confirm the observed anisotropic nature of lithium transport (**Table 2**). Indeed, *β*-LZC shows almost equal directional ionic conductivities, each *ca.* 0.1 S $cm^{-1}$. In contrast, for both ordered and disordered *α*-LZCs, ionic conductivities along *x* and *y* directions demonstrate similar contributions to the cumulative ionic conductivity values, while the directional conductivity along *z*-axis significantly dominates, accounting for more than 60% of collective conductivity. This enhances our understanding that *α*-LZCs have a stronger preference for lithium-ion transport along the interlayer direction compared to isotropic intralayer driven diffusion observed in *β*-LZC. The directional correlation effects summarized in **Table S4** and **Table S5** show that for *α*-LZC phases, the degree of correlated motion along the *z*-axis is essentially lower compared to the *x* and *y*-axis. Such difference indicates significant correlated contribution to the lithium motion, particularly in the interlayer direction. Meanwhile, the single-layer crystallographic arrangement of zirconium results higher *z*-direction ionic conductivity for disordered *α*-LZC than ordered one, which directly leads to higher overall ionic conductivity. For *β*-LZC, the degree of the correlated motion is negligible which is consistent with weak correlation effects and dominating contribution of lithium individual dynamics.

**Table 2.** Lithium-ion transport characteristics in $Li_2ZrCl_6$, LZC, expressed by cumulative, *total*, and decomposed, *directional*, *xyz*, ionic conductivities, $\sigma$ (S $cm^{-1}$), for ordered and disordered $\alpha$-LZC and $\beta$-LZC calculated at 950 K by means of deep learning-accelerated molecular dynamics (see **Table S4** for additional details).

| | $\sigma_x$ | $\sigma_y$ | $\sigma_z$ | $\sigma_{total}$ |
|---|---|---|---|---|
| ordered $\alpha$-LZC | 0.184 | 0.210 | 0.570 | 0.964 |
| disordered $\alpha$-LZC | 0.203 | 0.151 | 0.844 | 1.198 |
| $\beta$-LZC | 0.106 | 0.099 | 0.099 | 0.304 |

These quantitative insights are visually corroborated by examining the trajectories of arbitrary lithium ions (**Figure 7**). As can be clearly seen, in all LZC phases lithium ions exhibit the motion extending in multiple spatial directions, which confirms their three-dimensional diffusive behavior. Nevertheless, there are also significant qualitative differences among the phases. In $\beta$-LZC, lithium mobility is spatially restricted, with fewer displacements and shorter trajectories. At the same time, there is less motion observed in the *z* direction. In contrast, both ordered and disordered $\alpha$-LZCs show significantly more frequent and extensive lithium displacements, which as suggested above, leading to higher ionic conductivity. The pronounced anisotropic interlayer transport behavior in $\alpha$-LZCs, quantitatively indicated by the significant dominance of $z$-directed conductivity. Apart from this, another feature can be observed within the considered trajectories – lithium diffusion is not continuous but rather occurs through a sequence of discrete jumps – hopping events – between different coordination environments, separated by brief stationary periods. This intermittent hopping behavior is consistent with our quantitative results regarding the critical role of correlation effects in $\alpha$-LZCs, especially along the interlayer direction.

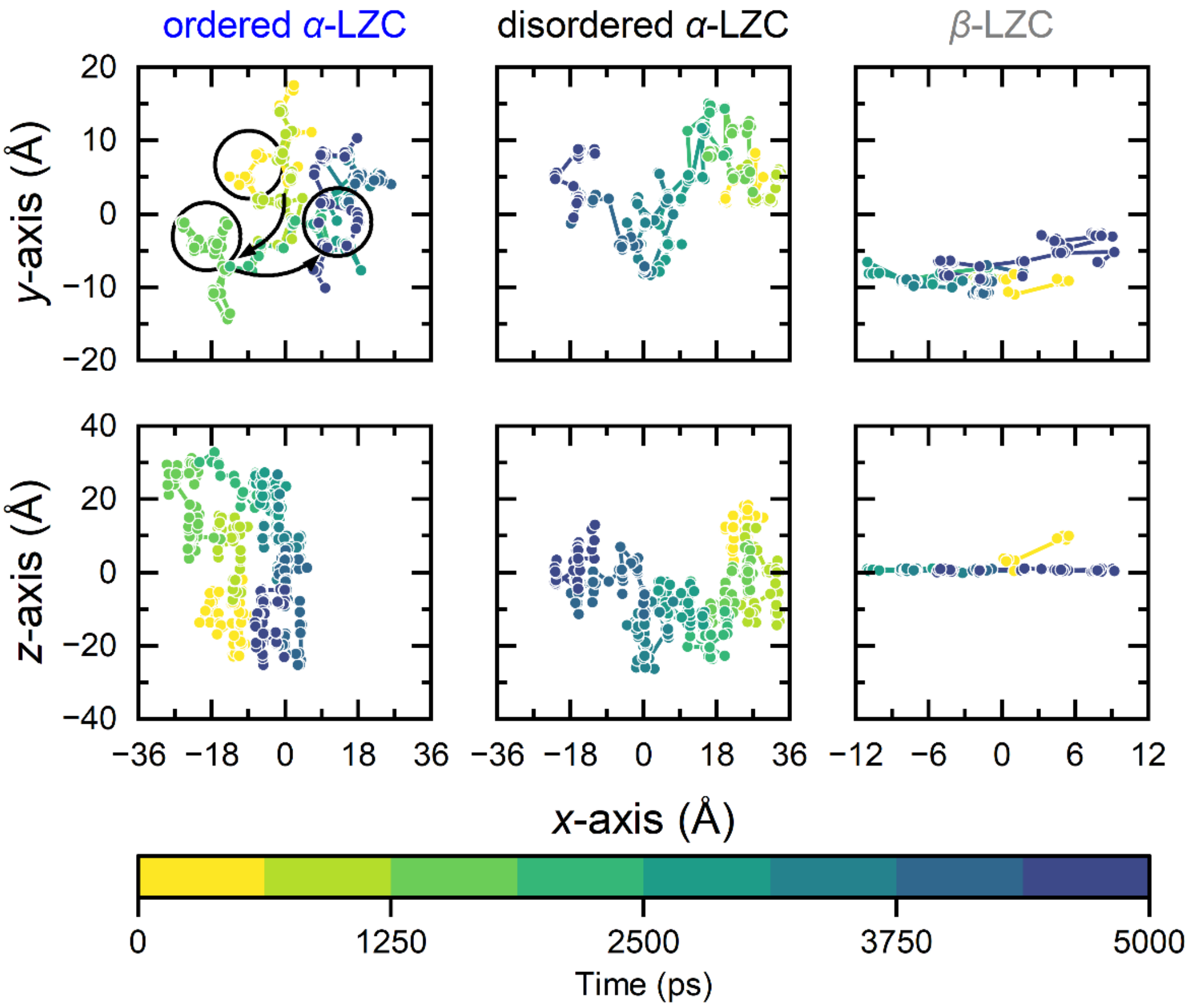

**Figure 7.** Lithium-ion migration pathway in $Li_2ZrCl_6$, LZC, for ordered *α*-LZC (*left column*), disordered *α*-LZC (*middle column*), and *β*-LZC (*right column*) phases, expressed by *xy*- (*top row*) and *xz*-plane (*bottom row*) projections to a single arbitrary lithium ion trajectory over 5 ns simulation time calculated at 950 K by means deep learning-accelerated molecular dynamics simulation. Each *circle* represents the position of lithium ion at the given time range. *Black arrows* stands for visual guidance and represent the hopping events between different sites.

To further investigate the dynamic sensitivity of lithium diffusion to its surrounding environment, we analyzed the individual-part of the van Hove correlation function (**Equation (10)**), which describes the time-dependent displacement distributions relative to the initial position of individual atoms. **Figure 8a** demonstrates that across all the considered LZC phases, the lithium-ion displacement probabilities follow a similar trend over time. At shorter time intervals, the shape of the distribution are notably skewed, while longer simulation times lead to essentially smoother distributions. This pattern highlights the discrete nature of lithium motion – hopping mechanism. At shorter distances, each curve is dominated by a sharp peak indicating that the position of the atoms remains predominantly at the initial lattice site and within original coordination environment. The characteristic secondary peaks observed in these distributions further confirm and quantify lithium hopping behavior. Particularly, upon increasing the distance, appearance of the new peak appears at *ca.* 3 Å indicates that lithium ions exhibit their first hopping event from the position at 1.8-2.5 Å to the adjacent lattice sites located at 5.3-5.6 Å. Such difference between the first and second minima (characteristic local ion jump-length) may indicate a partial trapping by the well-defined Zr···Cl lattice, with a pronounced sequence of ion jump from $\beta$-LZC (2.8 Å) to disordered $\alpha$-LZC (3.4 Å) and ordered $\alpha$-LZC (3.8 Å). In case $\alpha$-LZC phases this peak gradually broadens and reduces, while of its maximum shifts towards the 3.6 Å along the simulation time. This trend confirms the presence of structural constrains in LZC lattice with narrower channels in $\beta$-LZC phase and relatively flexible coordination in $\alpha$-LZC. On the other hand, the secondary peaks at larger distances *ca*. 7 and 10 Å further appear and growth with simulation time, corresponding to subsequent hopping events across multiple lattice positions. The increasing of the intensities of these peaks at longer simulation times indicates how fast the lithium ions escaping its coordination environment, where the faster site-to-site hopping observed in disordered $\alpha$-LZC, slower – in ordered $\alpha$-LZC and slowest in $\beta$-LZC. These integer multiples of the initial jump distance clearly reflect that lithium ions traverse longer spatial ranges through sequential hopping events, consistent with the *xy*- and *xz*-projection trajectories.

At the same time, precising the van Hove correlation function in terms of nearest neighboring lithium can further shed light on hopping behavior (**Figure 8b**). For $\alpha$-LZCs, the distributions exhibit more fragmented and discontinuous probability patterns. These shorter, more frequent bands appear rapidly and are dispersed across the full trajectory, indicating a dynamic landscape where lithium ions frequently hop between adjacent lattice sites. The discrete and temporally localized nature of these bands reflects a high frequency of hopping events and higher lithium diffusivity. At the same time, in $\beta$-LZC, the displacement probabilities appear longer, more continuous and horizontally extended overtime. This pattern reflects residence times within the lithium coordination environment and fewer discrete hopping events, hence, indicating that lithium ions in $\beta$-LZC tend to remain localized for longer periods before undergoing diffusion. Such behavior is a characteristic of systems with low ionic mobility and consistent with the reduced ionic conductivity illustrated above.

A more quantitative and straightforward relationship between lithium hopping and ionic conductivity is provided by the analysis of residence time (**Equation (11)-(13)**), which describes a duration of spatial confinement of lithium ion (**Table 3**). Among three studied LZC phases, disordered $\alpha$-LZC exhibits the shortest residence time, indicating a rapid hopping between lattice sites along ion pathway. On the other hand, $\beta$-LZC shows the longest residence time, around an order of magnitude larger compared to $\alpha$-LZCs. Microscopically, higher residence time implies that lithium ions spend more time confined to a single lattice site rather than moving through the structure. Such prolonged stationary periods directly affected in decreasing of lithium mobility and, hence, reducing ionic conductivity and increasing activation energy. Indeed, the analysis of the effective length scale of ion hopping dynamics (**Equation (14)**) reveal this behavior within the characteristics jump-length of 8.85 Å for $\beta$-LZC, 3.34 Å – for disordered $\alpha$-LZC and 3.44 Å – for ordered $\alpha$-LZC. While in $\beta$-LZC, lithium accumulates its motion though the repeated short-range oscillations, the short residence times observed in $\alpha$-LZCs promotes faster ion local rearrangements within geometric lattice-site separations as shown in **Figure 8**. Apart from this, residence time can also shed a light on the correlated diffusion phenomenon. By comparing the residence time of a single lithium ion in isolation with that when another atom occupies the adjacent site, we observed a significantly shorter residence time in the latter case. This suggests that lithium ions tend to avoid adjacent sites occupied by other lithium ions due to repulsive forces. Consequently, this unique effect significantly reduces local residence time, promoting rapid hopping events and enhancing ionic conductivity.

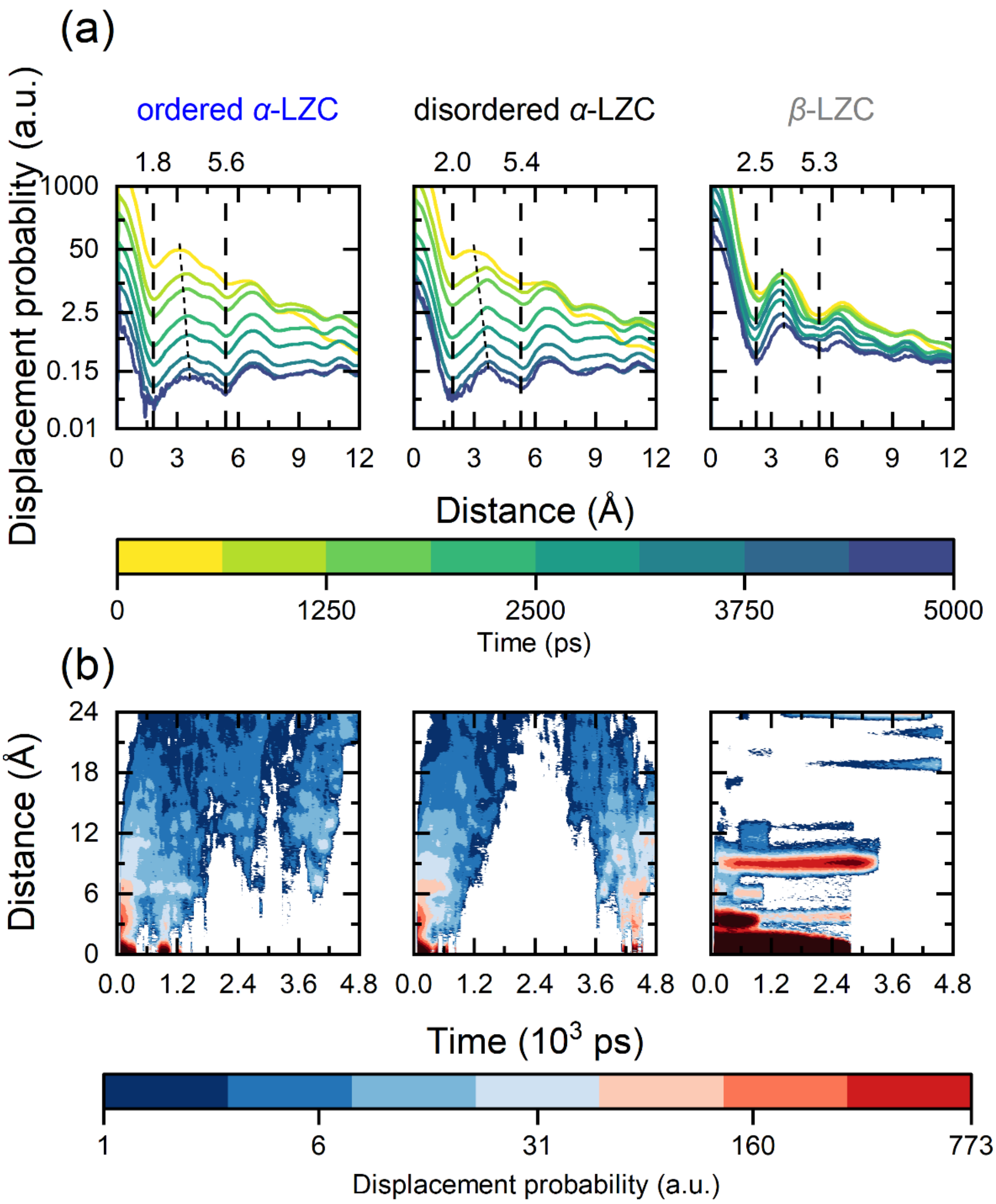

**Figure 8.** Lithium-ion transport mechanism in $Li_2ZrCl_6$, LZC, for ordered *α*-LZC (*left column*), disordered *α*-LZC (*middle column*) and *β*-LZC (*right column*) phases, expressed by (**a**) lithium-ion displacement probability represented by the individual-part of the van Hove correlation function along 5 ns trajectory. *Black dashed lines* between 1.8-2.5 Å and 5.3-5.6 Å stands for visual guidance and represent preferential hopping distance, while the *short-dashed line* additionally represents the evolution of the peak position for the first hopping even; (**b**) two-dimensional lithium-ion displacement probability density map with respect to its nearest neighbor calculated at 950 K by means deep learning-accelerated molecular dynamics simulation. Mind the logarithmic scale of displacement probability.

**Table 3.** Lithium-ion transport characteristics in $Li_2ZrCl_6$, LZC, expressed by the average residence time, $T_{res}$ (ps), in one site and to adjacent lithium (see **Figure S9** for corresponding autocorrelation function representations) for ordered and disordered $\alpha$-LZC and $\beta$-LZC calculated at 950 K by means of deep learning-accelerated molecular dynamics.

| | $T_{res}$ in one site | $T_{res}$ to adjacent Li |
|---|---|---|
| ordered $\alpha$-LZC | 5.488 | 0.697 |
| disordered $\alpha$-LZC | 5.241 | 0.584 |
| $\beta$-LZC | 87.226 | 15.422 |

To gain another spatially resolved understanding of lithium diffusion pathways in three-dimensional space, we calculated the spatial distribution function for each LZC structure (**Figure 9**). For facilitating observation, the spatial distribution is clipped to the original unit cell size of 1 × 1 × 1 (see **Figure S10** for the complete trajectory and original system size). In the visualizations, higher isosurface levels (darker and denser surfaces), representing the time-averaged probability density of lithium ions, correspond to the regions where lithium ions are most frequently located. These regions form highly localized, nearly spherical densities, reflecting the spatial confinement of lithium ions within well-defined coordination environments. Simultaneously, lower isosurface levels (lighter and sparser surfaces) capture regions or lower-probability associated with transient ion motion during hopping events between adjacent sites. These appear as channel-like features connecting high-density regions and directly visualize the pathways of lithium-ion migration throughout the material. Comparison between the three LZC systems reveals clear structural differences in the nature of diffusion pathways. In both ordered and disordered $\alpha$-LZCs, the channel-like isosurfaces are more prominent along the $z$-direction, with higher intermediate isosurface levels observed between high-occupancy sites. Oppositely, the isosurface values in the $x$- and $y$-axes are noticeably lower, indicating a reduced likelihood of hopping along those directions. This anisotropic channel connectivity directly supports our conclusions based on mean-squared displacement decomposition and van Hove analysis, which demonstrated that $\alpha$-LZCs favor interlayer diffusion along the $z$-axis. Moreover, these findings also align with the directional ionic conductivities and Haven's ratio analysis, where the $z$-direction consistently contributed more significantly to overall ion transport in $\alpha$-LZC phases. For $\beta$-LZC, the resulting distributions appear more spatially uniform, with less pronounced variation in isosurface levels across different spatial directions. This isotropic spatial distribution is consistent with the above observed equal contributions to ionic conductivity along $x$, $y$, and $z$-axes and with the reduced correlation effects, as indicated by a Haven's ratio close to unit. In addition, the sharp localization of lithium-ion density at high isosurface levels compared to the relatively low-density bridging regions highlights the intermittent nature of hopping transport. Thus, lithium ions spend the majority of their time at stable sites, while the time spent moving between these sites is negligible. This visual analysis further enhances the critical role of residence time in determining ionic conductivity: shorter residence times correlate with more frequent hopping and higher conductivity observed in $\alpha$-LZCs, whereas longer residence times suppress overall ion mobility in $\beta$-LZC.

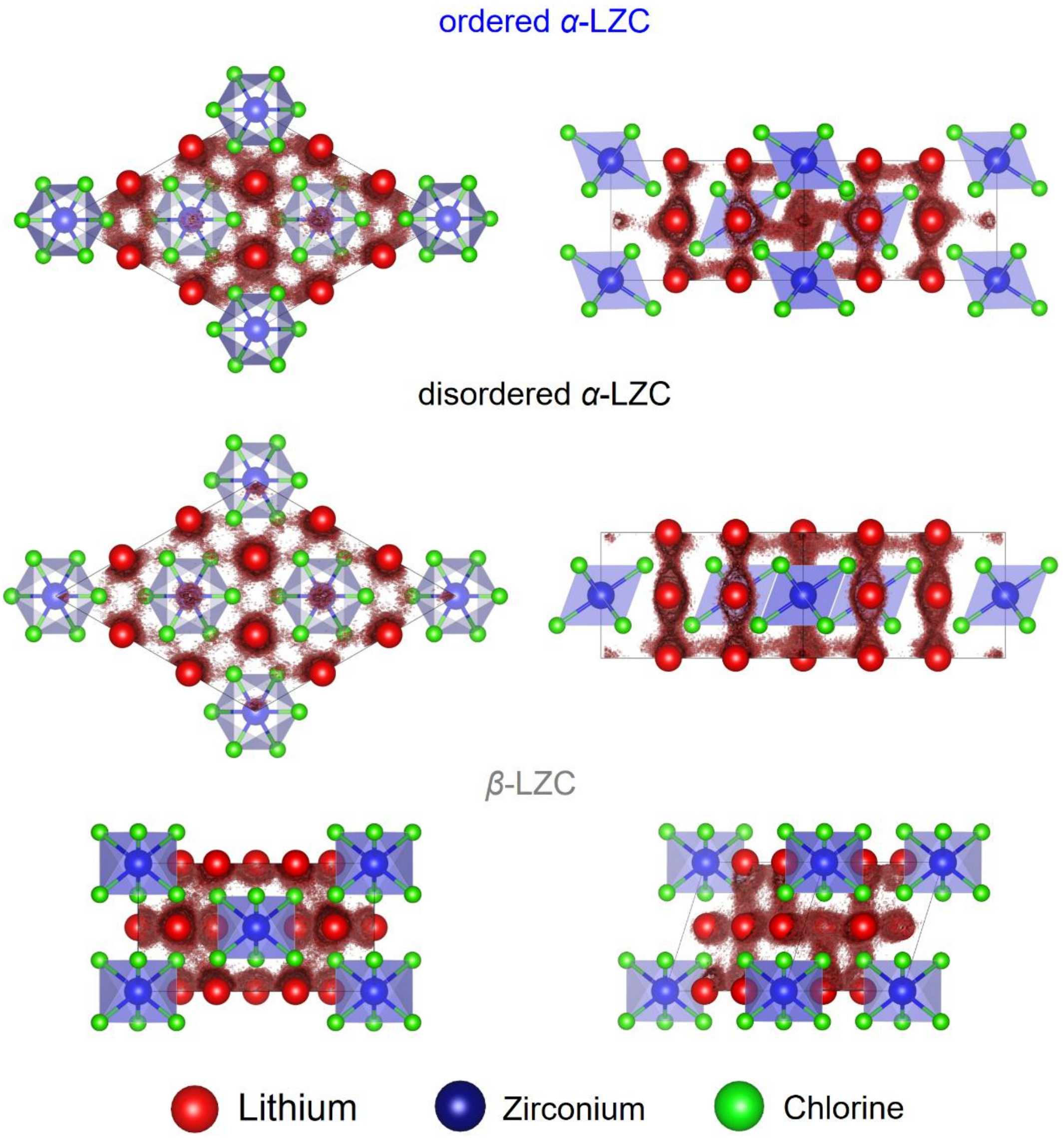

**Figure 9.** Lithium-ion conduction pathway in $Li_2ZrCl_6$, LZC, for ordered *α*-LZC (*top row*), disordered *α*-LZC (*middle row*) and *β*-LZC (*bottom column*) phases, expressed by spatial distribution function of lithium-ion, shown in top (*left column*) and side (*right column*) views calculated at 950 K by means deep learning-accelerated molecular dynamics simulation. Only the 1 × 1 × 1 unit cell matching the size of *ab initio* molecular dynamics supercell is presented for clarity (see **Figure S10** for large-scale representations).

To gain a deeper understanding of the local structure organization promoting the difference in lithium-ion diffusion in different LZC phases, we conducted an in-depth analysis of their respective structural features. First, we compared the X-ray diffraction patterns of the LZC structures along the DLP trajectory with previously obtained experimental measurements. The results presented in **Figure 10a** indicate that the LZC structures are consistent with the experimental observations, specifically, for the main peaks in the X-ray diffraction patterns of *α*-LZC (17-18°, 33°, 42°, 51°) and *β*-LZC (15-20°, 28-30°, 35°, 50°), the atomic arrangements in the simulated structures align closely with respect to each other. This not only confirms the high accuracy and descriptive capability of our DLP model for LZC but also confirms the long-range order of its atomic distribution.

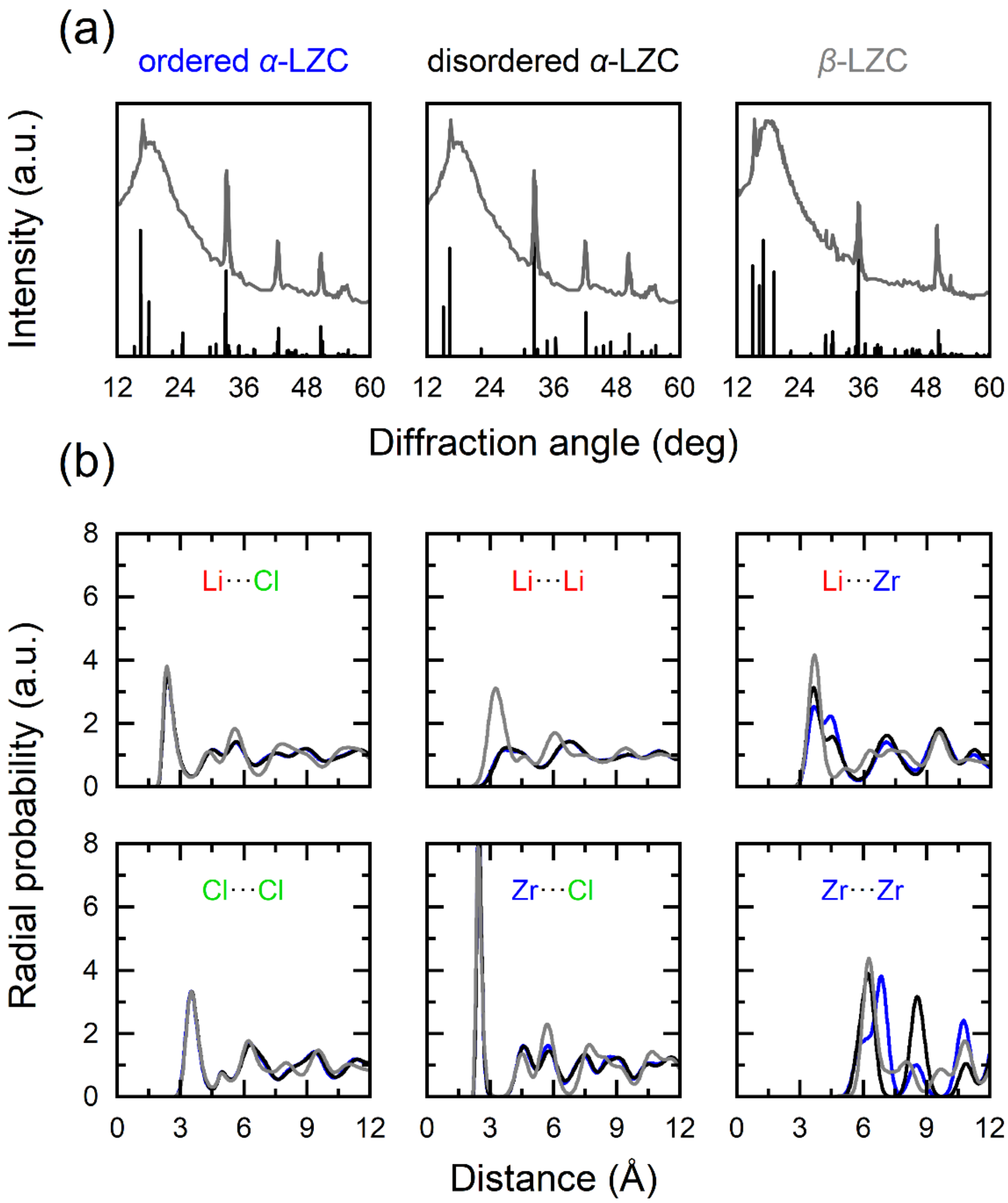

**Figure 10.** Structure characteristics in $Li_2ZrCl_6$, LZC, for ordered α-LZC (*left column*), disordered α-LZC (*middle column*) and β-LZC (*right column*) phases, expressed by (**a**) the comparative analysis of X-ray diffraction patterns, where the *grey* lines represent the experimentally measured X-ray diffraction [14], while the *black* corresponds to the results obtained at 950 K by means of deep learning-accelerated molecular dynamics simulation; (**b**) radial distribution functions between lithium, zirconium and chlorine atomic pair combinations (Li···Cl, Li···Li, Li···Zr, Cl···Cl, Zr···Cl and Zr···Zr) calculated for ordered α-LZC (*blue line*), disordered α-LZC (*black line*) and β-LZC (*grey line*) phases.

Building on the structural validation, we further explored the local atomic arrangements using radial distribution functions for all atomic pairs (**Figure 10b**). This approach provides detailed insights into the short-range radial structural features of different LZC phases, complementing the long-range order revealed by the X-ray diffraction analysis. The obtained results show the multiple peaks at shorter separation distances across all distributions, indicating that the atomic arrangement in LZC is relatively compact and well-ordered at the local scale. Notably, for all atomic pairs involving chlorine atoms (Li···Cl, Cl···Cl and Zr···Cl), the radial distribution profiles for the three structural phases of LZC are almost identical. This similarity suggests that chlorine atoms play a structurally stabilizing but passive role, contributing in the same manner across different phases without strongly differentiating between them. This observation is also consistent with direct visualization of atomic configurations (**Figure 2**), where chlorine atoms are consistently arranged around zirconium in all LZC structures, thereby forming $ZrCl_6^-$ octahedral units. The particularly sharp and prominent Zr···Cl distributions at 2.512 Å confirm the strong local coordination and bonding between zirconium and chlorine atoms, which is consistent with octahedral geometry. On the other hand, the Zr···Cl distribution shows broader and less intensive peaks corresponding to the weaker and more variable electrostatic interactions between lithium and nearest chlorine atoms. For Li···Li, Li···Zr and Zr···Zr atomic pairs, the radial distributions exhibit the similar first peak positions at 3.632 (Li···Li), 3.632 (Li···Zr) and 6.224 Å (Zr···Zr), respectively. This consistency in the peak position indicates that the overall crystal lattice geometry is conserved for all LZCs. Thus, the difference in lithium transport behavior among the different structures is not primarily due to large-scale structural distortions but instead arises from subtle changes in site connectivity and dynamic correlation.

Another noticeable feature observed in ordered $\alpha$-LZC is that several distributions (Li···Li, Li···Zr, and Zr···Zr) exhibit split first peaks with similar interaction probabilities. These split peaks may be related to the layered and more periodically modulated structure of the ordered $\alpha$-LZC phase. The appearance of an additional peak approximately 1 Å beyond the primary peak reflects interlayer interactions that are not symmetrically equivalent, which is a hallmark of the denser layered stacking unique to the ordered $\alpha$-LZC phase. The absence of split features in disordered $\alpha$-LZC and $\beta$-LZC suggests stability and symmetry in the interatomic distances between neighboring layers along the $z$-axis direction. This denser layered stacking observed in ordered $\alpha$-LZC, despite sharing the same space group as disordered $\alpha$-LZC, likely contributes to their differing ionic conductivities and activation energies. The asymmetry between the Zr···Zr and Li···Zr layers in ordered $\alpha$-LZC introduces additional barriers to lithium diffusion across these layers. This asymmetry caused by particular zirconium crystallographic arrangements result in a marginally reduced ionic conductivity for ordered $\alpha$-LZC relative to disordered $\alpha$-LZC.

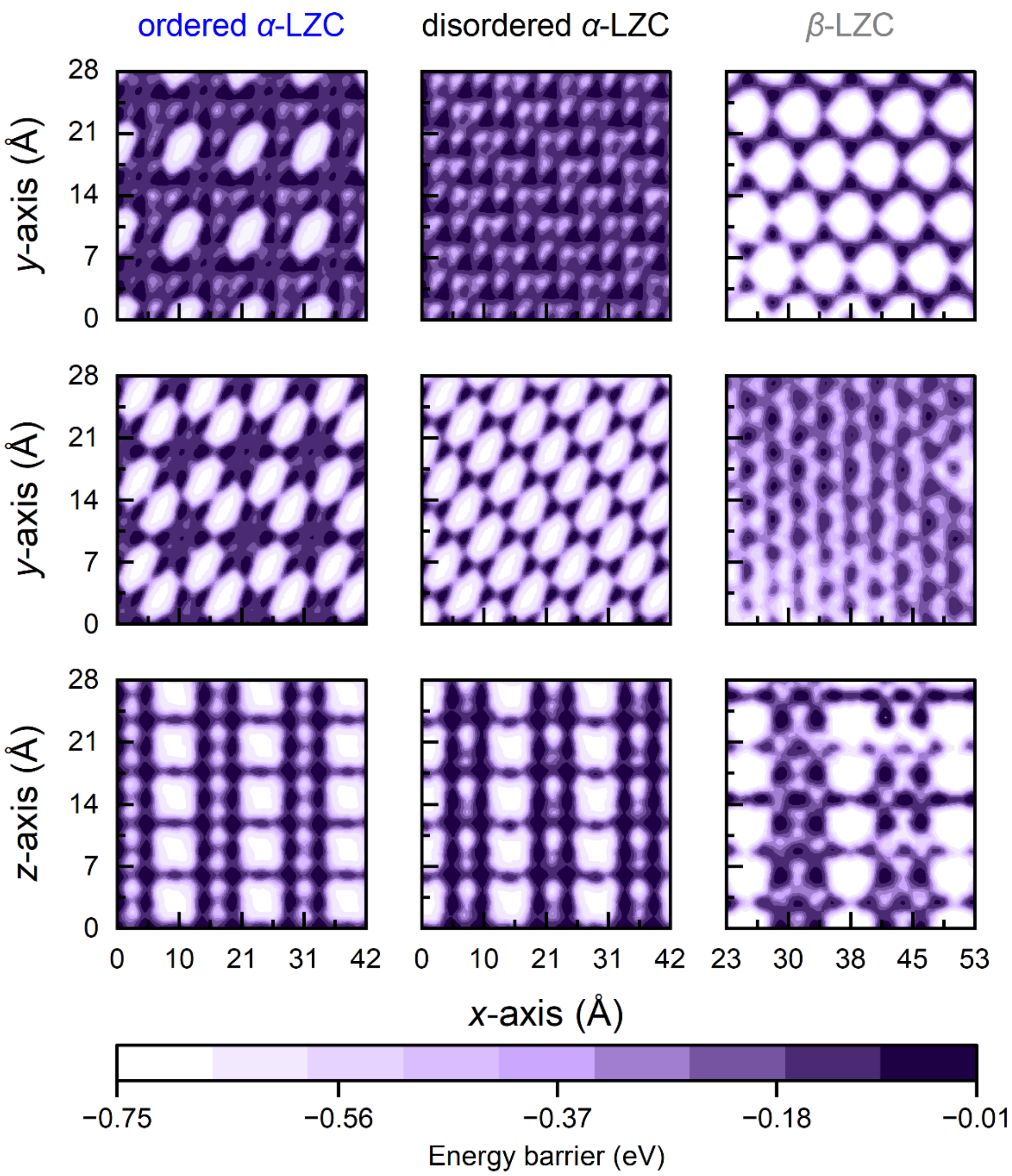

**Figure 11.** Energetics of lithium-ion migration pathway in $Li_2ZrCl_6$, LZC, for ordered *α*-LZC (*left column*), disordered *α*-LZC (*middle column*) and *β*-LZC (*right column*) phases, expressed by edge/bottom *xy*- (*top row*), intermediate *xy*- (*middle row*) and *xz*-plane (*bottom row*) contour projections of potential of mean force calculated at 950 K by means deep learning-based molecular dynamics simulation. The energy gap between different paths represents the energy barrier that a single lithium needs to cross when moving.

Despite a comprehensive analysis of lithium-ion dynamics and local structure organization, the above presented analysis still lacks a direct description of the energetic pattern that governs the variation of ionic conductivity in studied LZC phases. To fulfill this gap, we performed potential of mean force analysis to elucidate the Gibbs free energy landscape of the lithium motion. The resulting projections potential of mean force for the three LZC structures (**Figure 11**) reveal energy minima corresponding to stable lithium sites and diffusion barriers with energy increasing, ranging from *ca*. −2.38 eV to −0.04 eV at 950 K (between −0.75 and −0.01 eV at room temperature, respectively). These projections helps to define where lithium ions are encounter the hindrances – barriers – determined by the features of the spatial organization of LZC. Notably, due to the difference in zirconium distribution across the phases, these maps contain slices taken at two different *z*-axis positions in the *xy*-plane: the edge/bottom layer (*z*-fractional coordinate = 0.0) and the intermediate layer (*z*-fractional coordinate = 0.5). This layered difference based on various zirconium localization density highlights how the zirconium arrangement modulates lithium mobility differently at different depths, with consequent changes in the intralayer potential barrier.

Indeed, for both ordered and disordered *α*-LZC phases, the projections in the *xy*-plane show that the energy barrier for intralayer diffusion is relatively high, characterized by extended dark regions separating with white minima. However, this intralayer steric hindrance varies between the layers, particularly in the quantitative analysis of the energy barriers: in the edge/bottom *xy*-layers, the lower localization density of zirconium distribution leads to more continuous low-energy pathways. More precisely, the zirconium-free *xy*-plane results in an interlayer barrier of only 0.459 eV and 0.535 eV for disordered and ordered *α*-LZC (0.145 and 0.169 eV at room temperature). By comparison, the intermediate *xy*-plane exhibits a pronounced high-energy barrier (whiter, more fragmented pattern) due to the higher zirconium localization density, trapping lithium ions in isolated minima and producing a cage-like confinement – the intralayer barriers of disordered and ordered *α*-LZC in the intermediate *xy*-plane reach 0.801 eV and 0.744 eV (0.253 and 0.235 eV at room temperature). Meanwhile, the *xz*-plane exhibits the lowest barrier along the *z*-direction, indicating a preferential pathway for interlayer lithium transitions – the interlayer barriers for disordered and ordered *α*-LZC are only 0.254 eV and 0.336 eV (0.080 and 0.106 eV at room temperature). These lowest interlayer barriers (smaller than 59.4% compared with intralayer one) further verify interlayer motion tendency of lithium ions, which is facilitated by channel-like connectivity along the *z*-axis. It also confirms that the asymmetry, caused by compact arrangement of zirconium, between the Zr···Zr layer and the Li···Zr layer in the ordered *α*-LZC does introduce additional energy barriers for the diffusion of lithium compared with disordered one. This gap in energy barriers, according to the Arrhenius relation, also verifies the faster interlayer hopping rate in dynamics: the interlayer rates of disordered and ordered *α*-LZC are 12.36 times and 11.44 times faster than intralayer one, respectively.

Considering *β*-LZC, the energy landscape exhibits a more intralayer-driven isotropic distribution, with interconnected low-energy pathways in the *xy*-plane and a wider high-energy region in the *xz*-plane. Here, the layer distinction is clearly seen: the edge/bottom *xy*-layer shows slightly broader high-energy regions influenced by surface-like effects. The intermediate *xy*-layer features more uniform minima that enhance connectivity for intralayer migration, with lower barriers due to reduced zirconium density variations. The calculated energy barriers reflect this change, where the interlayer barrier is 1.628 eV (0.514 eV at room temperature), while the intralayer barriers are 1.564 eV in the bottom layer and 1.498 eV in the intermediate layer (0.494 and 0.473 eV at room temperature). These results indicate that intralayer migration is the main mechanism in *β*-LZC, particularly favored in the intermediate layer. At the same time, based on the small difference between the interlayer and intralayer barriers (only 8.7%), *β*-LZC does not have the same strong single-transition tendency as *α*-LZC.

To further confirm the energy barrier differences revealed by the potential of mean force analysis, we carried out further electronic structure and bonding analyses. First, the partial density of states of zirconium reveals that the disordered $\alpha$-LZC exhibits the highest density of occupied zirconium states near the Fermi level, whereas $\beta$-LZC shows the lowest (**Figure S11**). This indicates that the zirconium electronic states in the disordered $\alpha$-LZC are less stabilized compared with those in the ordered $\alpha$-LZC and $\beta$-LZC, thereby weakening the zirconium framework stability and facilitating the relative motion of lithium ions with respect to zirconium. Second, the analysis of the negative integrated crystal-orbital Hamilton population of zirconium–chloride bonds, which serves as a measure of bonding strength within the $ZrCl_6^{2-}$ octahedra, shows that: disordered $\alpha$-LZC > ordered $\alpha$-LZC > $\beta$-LZC (**Figure S12**). This confirms that disordered $\alpha$-LZC possesses the weakest zirconium–chloride covalent interactions and consequently a more softened $ZrCl_6^{2-}$ octahedral framework compared to the other two phases. Consistently, the average zirconium–chloride bond length analysis supports this observation: disordered $\alpha$-LZC exhibits a longer zirconium–chloride bond (2.483 Å) than ordered $\alpha$-LZC (2.481 Å) and $\beta$-LZC (2.481 Å), further validating the weakened zirconium–chloride bonding in the disordered structure. Thereby, these results provide a coherent explanation for the lower energy barrier of disordered $\alpha$-LZC: the loosely packed zirconium sublattice reduces the electronic stability of zirconium states, hence producing a softer $ZrCl_6^{2-}$ octahedra that facilitates lithium migration. These findings highlight how structural differences – particularly the degree of zirconium distribution disorder – induce subtle yet profound modifications in the electronic structure, which in turn regulate the lithium-ion transport by modulating the underlying Gibbs free-energy landscape.

## Conclusions

In the present work, in order to elucidate the underlying principles governing lithium-ion transport in $Li_2ZrCl_6$ (LZC) solid electrolyte, we conducted a comprehensive theoretical investigation. Recognizing the limitations of conventional theoretical approaches in studying electronic structure and dynamics in periodic systems, particularly their constraints on spatial and temporal scales, we have developed a novel deep learning-accelerated potential model tailored for the trigonal $\alpha$- and monoclinic $\beta$-LZC phases. This innovative approach bridges the gap between *ab initio* accuracy and computational scalability, enabling high-precision large-scale molecular dynamics simulations. In the framework of molecular dynamics simulations, our results reveal the lithium-ion transport mechanisms in LZCs, identifying individual and correlated diffusion events as key determinants beyond ionic conductivity. Among the three considered LZC phases – ordered $\alpha$-LZC, disordered $\alpha$-LZC and $\beta$-LZC – disordered $\alpha$-LZC exhibits the highest ionic conductivity. Examination of spatial ion dynamics illustrates that lithium in $\alpha$-LZCs not only exhibit individual, self, diffusion, but also shows the significant correlated effects between the lithium ions, especially in the interlayer pathway. On the other hand, ions in $\beta$-LZC exhibit isotropic translations and individual diffusion dominated by intralayer migration. The estimated degree of correlated ion motion confirms the critical role of correlated diffusion within $\alpha$-LZC and the dominance of individual diffusion across $\beta$-LZC. These observations were related to the ion migration mechanism through the stationary discrete states along its pathway *via* hopping mechanism. The analysis of the local structure organizations of the LZC phases confirms that the softening of $ZrCl_6^{2-}$ octahedra induced by unique zirconium crystal arrangement is the reasons for the different energy barriers and, in turn, the differences in dynamic behaviors. These insights not only enhance the understanding of lithium-ion transport mechanisms within LZCs, but also promote laying a robust and predictive computational framework capable of guiding the rational design of next-generation ion-conducting materials. By combining atomistic simulations with data-driven deep learning modeling, this framework provides a versatile platform to systematically explore compositional, structural, and dynamic factors influencing ionic conductivity. In future studies, this approach could be extended to a broader class of solid electrolytes, including sulfide-, oxide-, and halide-based systems, thereby accelerating the discovery and optimization of advanced materials tailored for high-performance, safe, and scalable energy storage technologies. Moreover, coupling this framework with experimental validation and machine learning could further refine predictive accuracy, paving the way for a new paradigm in solid-state electrolyte design.

**Supporting Information**

Description of deep learning potential training details, Structural parameters and convergence tests, Validation of model accuracy and stability, Energy and structural characteristics from simulations, Lithium ion transport properties and diffusion analysis, and Electronic structure and bonding characteristics

**Acknowledgments**

This research is supported by the Vehicle Technologies Office (VTO), Department of Energy (DOE), USA, through the Battery Materials Research (BMR) program. Argonne National Laboratory is operated for DOE by UChicago Argonne, LLC under the contract number DE-AC02-06CH11357. The authors gratefully acknowledge the computing resources provided on Bebop, a high-performance computing cluster, operated by the Laboratory Computing Resource Center at Argonne National Laboratory.

# Supporting Information

## Unveiling the Lithium-Ion Transport Mechanism in $Li_2ZrCl_6$ Solid-State Electrolyte *via* Deep Learning-Accelerated Molecular Dynamics Simulations

Hanzeng Guo[1,2], Volodymyr Koverga[1,2], Selva Chandrasekaran Selvaraj[1,2] and Anh T. Ngo[1,2]*

[1] Department of Chemical Engineering, University of Illinois Chicago, Chicago, IL 60608, United States

[2] Materials Science Division, Argonne National Laboratory, Lemont, IL 60439, United States

*Corresponding author's email: *anhngo@uic.edu*

### Section S1. Deep learning potential training theory

When constructing the deep learning potential (DLP), the local coordinate matrix $\mathcal{R}$, and local atomic environment matrix $\mathcal{R}^i \in \mathbb{R}^{N_i \times 3}$, are utilized, as illustrated in **Equations (S1)** and **(S2)**:

$$\mathcal{R} = \{\boldsymbol{r}_1^T, \dots, \boldsymbol{r}_i^T, \dots, \boldsymbol{r}_N^T\}^T, \boldsymbol{r}_i = (x_i, y_i, z_i) \tag{S1}$$

where $\boldsymbol{r}_i = (x_i, y_i, z_i)$ represents the Cartesian coordinates of atom $i$, and $N$ denotes the total number of atoms. The matrix $\mathcal{R}$ can be transformed into local environment matrices as:

$$\mathcal{R}^i = \{\boldsymbol{r}_{1i}{}^T, \dots \boldsymbol{r}_{ji}{}^T, \dots, \boldsymbol{r}_{N_i,i}{}^T\}^T, \boldsymbol{r}_{ji} = (x_{ji}, y_{ji}, z_{ji}) \tag{S2}$$

here $\boldsymbol{r}_{ij}^T \equiv \boldsymbol{r}_i^T - \boldsymbol{r}_j^T$ is defined as relative coordinates, whereas $j$ and $N_i$ are indexes and number of neighbors of $i^{\text{th}}$ atom within cutoff radius $\boldsymbol{r}_c$, respectively.

For constructing the sub-network, local atomic environment matrix $\mathcal{R}^i \in \mathbb{R}^{N_i \times 3}$ was mapped onto generalized coordinates $\tilde{\mathcal{R}}^{\mathrm{i}} \in \mathbb{R}^{N_i \times 4}$ by considering smooth cutoff parameter $r_{cs}$ [54]. In $\tilde{\mathcal{R}}^i$, $\boldsymbol{r}_{ji}$ is transferred to $\boldsymbol{r}_{ji} = \{s(\boldsymbol{r}_{ji}), \hat{x}_{ji}, \hat{y}_{ji}, \hat{z}_{ji}\}$. $s(\boldsymbol{r}_{ji})$ is a continuous and differentiable scalar weighting function applied to each component. By applying a smooth cutoff parameter $r_{cs}$, the components in $\tilde{\mathcal{R}}^i$ are allowed to smoothly go to zero at the boundaries of the local region defined by $r_c$. The weighting function $s(\boldsymbol{r}_{ji})$ reduces the weight of particles that are farther away from atom $i$. In addition, it removes the discontinuity introduced by the cutoff radius $r_{\mathrm{c}}$ from the DeepPot-SE model.

An embedding neural network $G(s(\boldsymbol{r}_{ji}))$ with three layers, each containing 20, 40, and 80 neurons, was then used to convert generalized coordinates $\tilde{\mathcal{R}}^{\mathrm{i}} \in \mathbb{R}^{N_i \times 4}$ to encoded feature matrix $\mathcal{D}^i$ can be written as:

$$\mathcal{D}^i = (\mathcal{G}^{i1})^T \, \tilde{\mathcal{R}}^i \, (\tilde{\mathcal{R}}^i)^T \, (\mathcal{G}^{i2}) \tag{S3}$$

where $\mathcal{G}^i$ is the local embedding matrix form $G(s(\boldsymbol{r}_{ji}))$ [54]. The incremental design of neuron counts was intended to progressively enhance the capability of network to extract features from local atomic environments. The initial layer, comprising 20 neurons, captured fundamental local geometric information. Subsequent layers progressively construct more complex nonlinear representations to characterize the diversity of interatomic interactions. This progressive structure ensured robust representation capability while maintaining control over the computational model complexity and mitigating the risk of overfitting. Encoded feature matrix $\mathcal{D}^i$ was set as descriptors input into a fitting neural network $\mathcal{F}_0$ with three layers, each comprising 120 neurons, that maps descriptors to atomic energies $E_i$. The configuration of 120 neurons per layer was designed to provide sufficient model capacity to accurately fit the complex PES while ensuring computational efficiency. The choice of a relatively large number of neurons satisfied the need for high-precision atomic energy predictions, particularly for long-range interactions and many-body effects. Total energy $E$, force $F$ and virial tensor $\Xi$ were calculated with **Equation (S4)-(S6)** [54,55]:

$$E = \sum_{i=0}^{i=N} E_i = \sum_{i=0}^{i=N} \mathcal{F}_0(\mathcal{D}^i) \tag{S4}$$

$$F_{i,\alpha} = -\frac{\partial \mathrm{E}}{\partial \boldsymbol{r}_{i,\alpha}} \tag{S5}$$

$$\Xi_{\alpha\beta} = -\sum_{\gamma} \frac{\partial \mathrm{E}}{\partial h_{\gamma\alpha}} h_{\gamma\beta} \tag{S6}$$

where $\boldsymbol{r}_{i,\alpha}$ and $F_{i,\alpha}$ denote the $\alpha^{\text{th}}$ component of the coordinate and force of atom $i$, respectively. $h_{\alpha\beta}$ represents the $\beta^{\text{th}}$ component of the $\alpha^{\text{th}}$ basis vector of the simulation region. The hyperbolic tangent activation function was applied in neural network to introduce nonlinearity and effectively train the complex atomic descriptor data $\mathcal{D}^i$.

**Table S1.** Lattice parameters comparisons of $Li_2ZrCl_6$, LZC, expressed by different van der Waals, vdW, interactions optimization. Comparisons involve lattice constants *a*, *b*, *c* (Å); lattice angles *α*, *β*, *γ* (°) and root mean square deviation, RMSD (Å).

| vdW interaction | *a* | *b* | *c* | *α* | *β* | *γ* | RMSD |
|---|---|---|---|---|---|---|---|
| | | | ordered *α*-LZC | | | | |
| Experiment | 10.971 | 10.971 | 5.931 | 90.000 | 90.000 | 120.000 | |
| No vdW | 11.056 | 11.056 | 6.062 | 90.452 | 89.548 | 120.480 | 0.164 |
| DFT-D2 | 10.874 | 10.874 | 5.854 | 89.934 | 90.066 | 120.422 | 0.171 |
| DFT-D3 | 10.827 | 10.827 | 5.819 | 89.970 | 89.970 | 120.387 | 0.185 |
| | | | disordered *α*-LZC | | | | |
| Experiment | 10.971 | 10.971 | 5.931 | 90.000 | 90.000 | 120.000 | |
| No vdW | 11.119 | 11.119 | 6.575 | 90.000 | 90.000 | 120.000 | 3.037 |
| DFT-D2 | 10.879 | 10.879 | 5.980 | 90.000 | 90.000 | 120.000 | 2.971 |
| DFT-D3 | 10.875 | 10.875 | 5.863 | 90.000 | 90.000 | 120.000 | 2.970 |
| | | | *β*-LZC | | | | |
| Experiment | 6.395 | 11.047 | 6.296 | 90.000 | 109.899 | 90.000 | |
| No vdW | 6.436 | 11.109 | 6.866 | 90.000 | 108.622 | 90.000 | 0.338 |
| DFT-D2 | 6.305 | 10.903 | 6.363 | 90.000 | 109.620 | 90.000 | 0.099 |
| DFT-D3 | 6.287 | 10.892 | 6.280 | 90.000 | 109.966 | 90.000 | 0.105 |

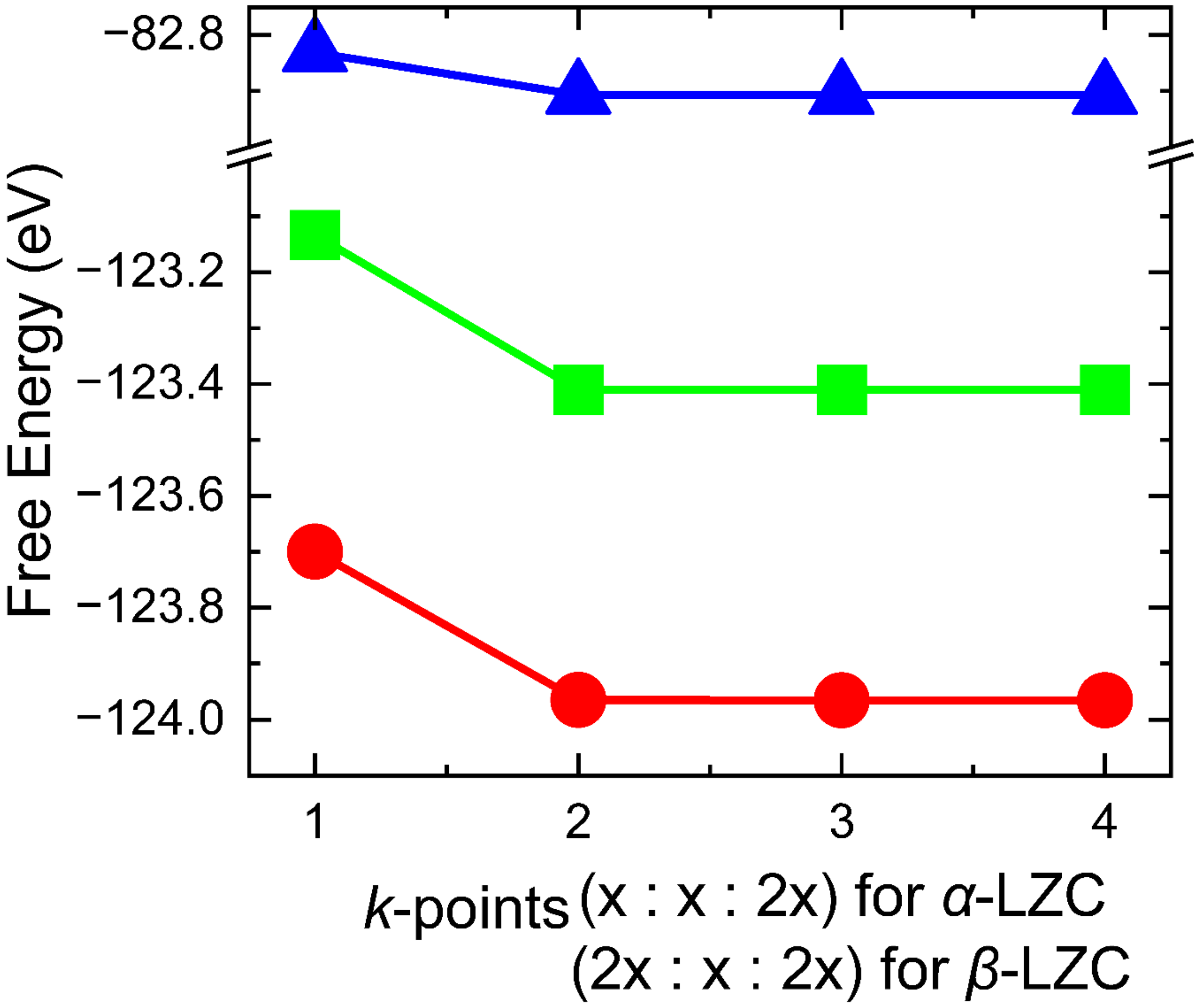

**Figure S1.** *k*-points test to $Li_2ZrCl_6$, LZC. The free energy converges when x⩾2, indicating that *k*-points with x⩾2 are sufficient for stable and accurate Density Functional Theory predictions. *Green points* represent ordered *α*-LZC, *red* – disordered *α*-LZC, *blue* – *β*-LZC.

**Table S2.** Statistical performance metrics of the trained deep learning potential for $Li_2ZrCl_6$, LZC, expressed by mean absolute error, MAE, and root mean square error, RMSE, for the loss function components – energy ($10^3$ eV atom$^{-1}$), force ($10^2$ eV Å$^{-1}$), and virial tensor ($10^2$ eV atom$^{-1}$).

| | ordered *α*-LZC | disordered *α*-LZC | *β*-LZC |
|---|---|---|---|
| Energy MAE | 3.158 | 2.005 | 4.728 |
| Energy RMSE | 5.261 | 3.301 | 6.626 |
| Force MAE | 4.292 | 4.755 | 4.864 |
| Force RMSE | 7.063 | 7.462 | 7.442 |
| Virial MAE | 1.166 | 1.031 | 1.400 |
| Virial RMSE | 1.677 | 1.575 | 2.204 |

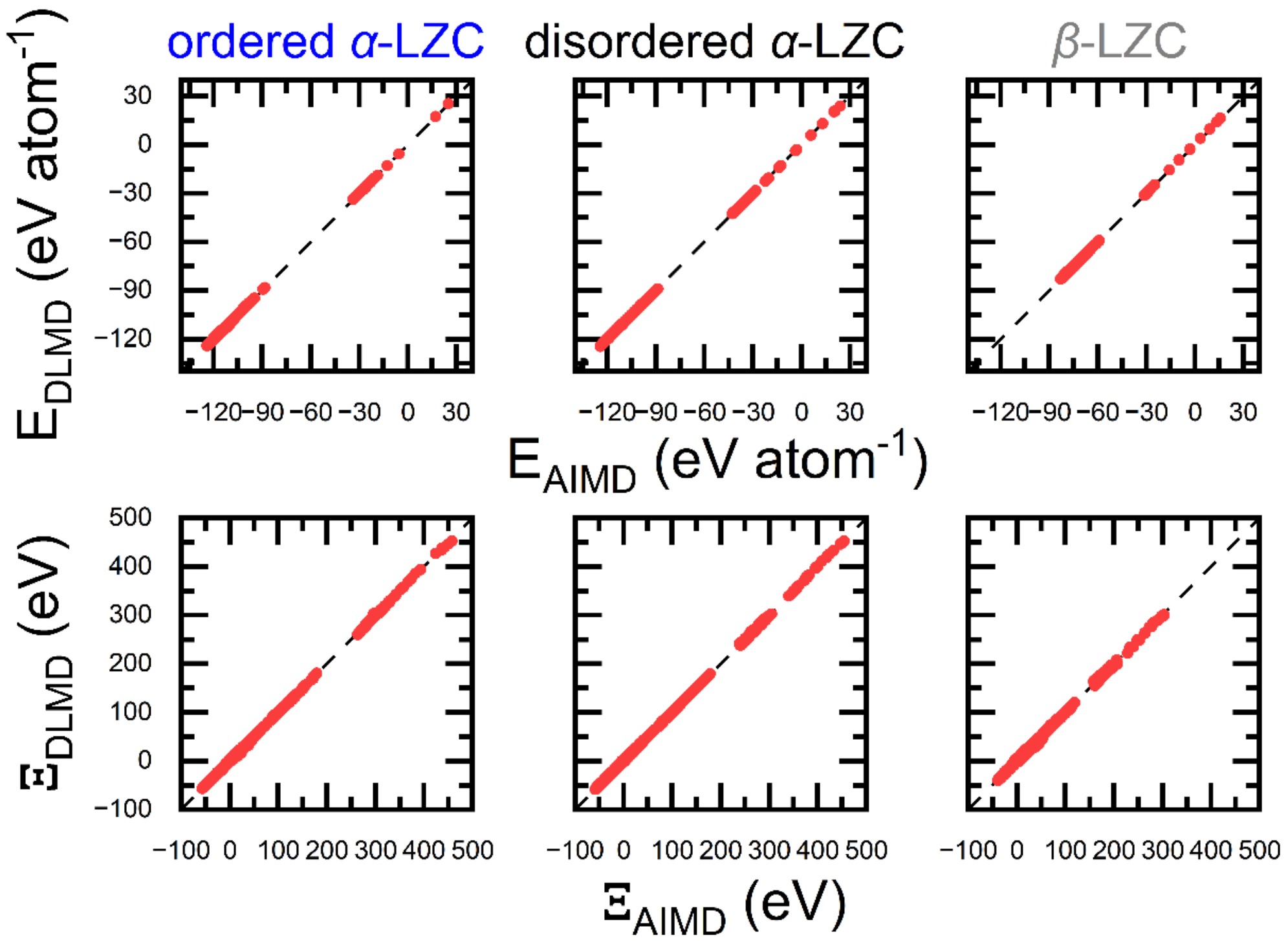

**Figure S2.** Validity of deep learning model for $Li_2ZrCl_6$, LZC, represented by correlation between energy, $E$ (*top row*) and virial tensor, Ξ (*bottom* row) predicted by deep learning potential, DLP and reference *ab initio* molecular dynamics simulation in ordered *α*-LZC (*left column*), disordered *α*-LZC (*middle column*) and *β*-LZC (*right column*) phases. In all cases, the estimated determination coefficient, $R^2$ = 1.00, indicates that 100% of the variance in *ab initio* $E$ and Ξ accurately captured by the DLP model within the range of accuracy considered. Black dashed line visual stands for visual guidance and represent ideal correlation with $R^2$ = 1.

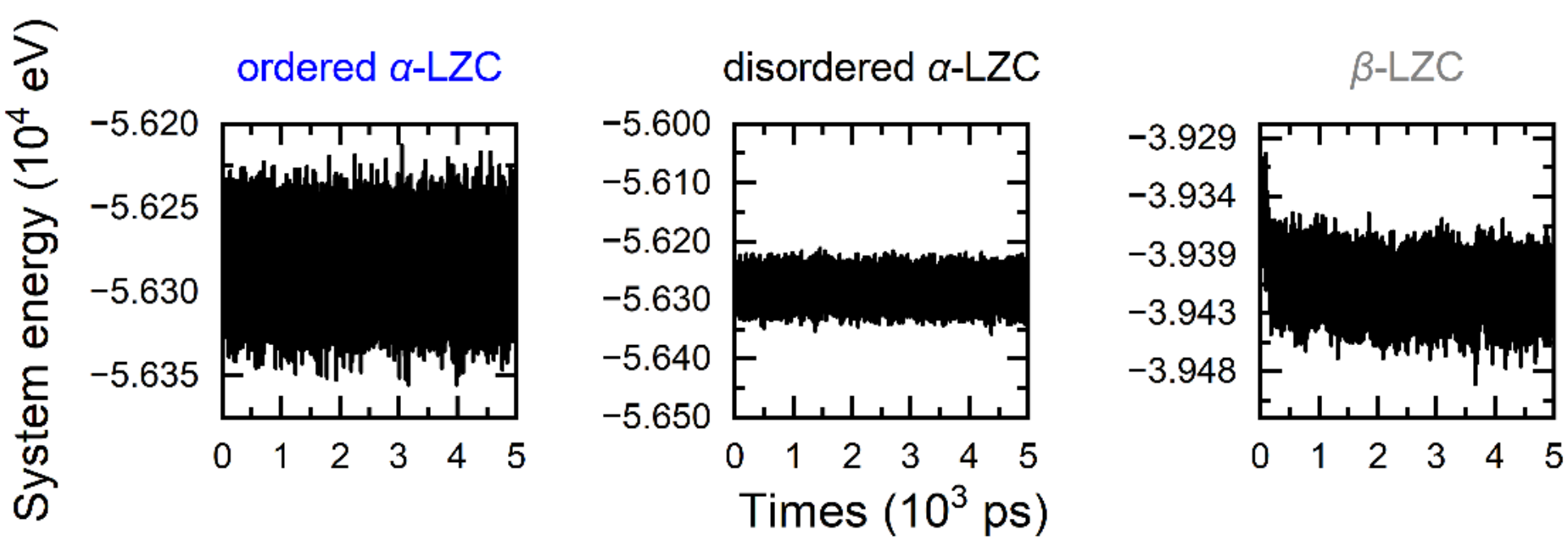

**Figure S3.** Energy variation of deep learning molecular dynamic simulations at 950K in $Li_2ZrCl_6$, LZC, for ordered *α*-LZC (*left column*), disordered *α*-LZC (*middle column*) and *β*-LZC (*right column*) phases during 5ns.

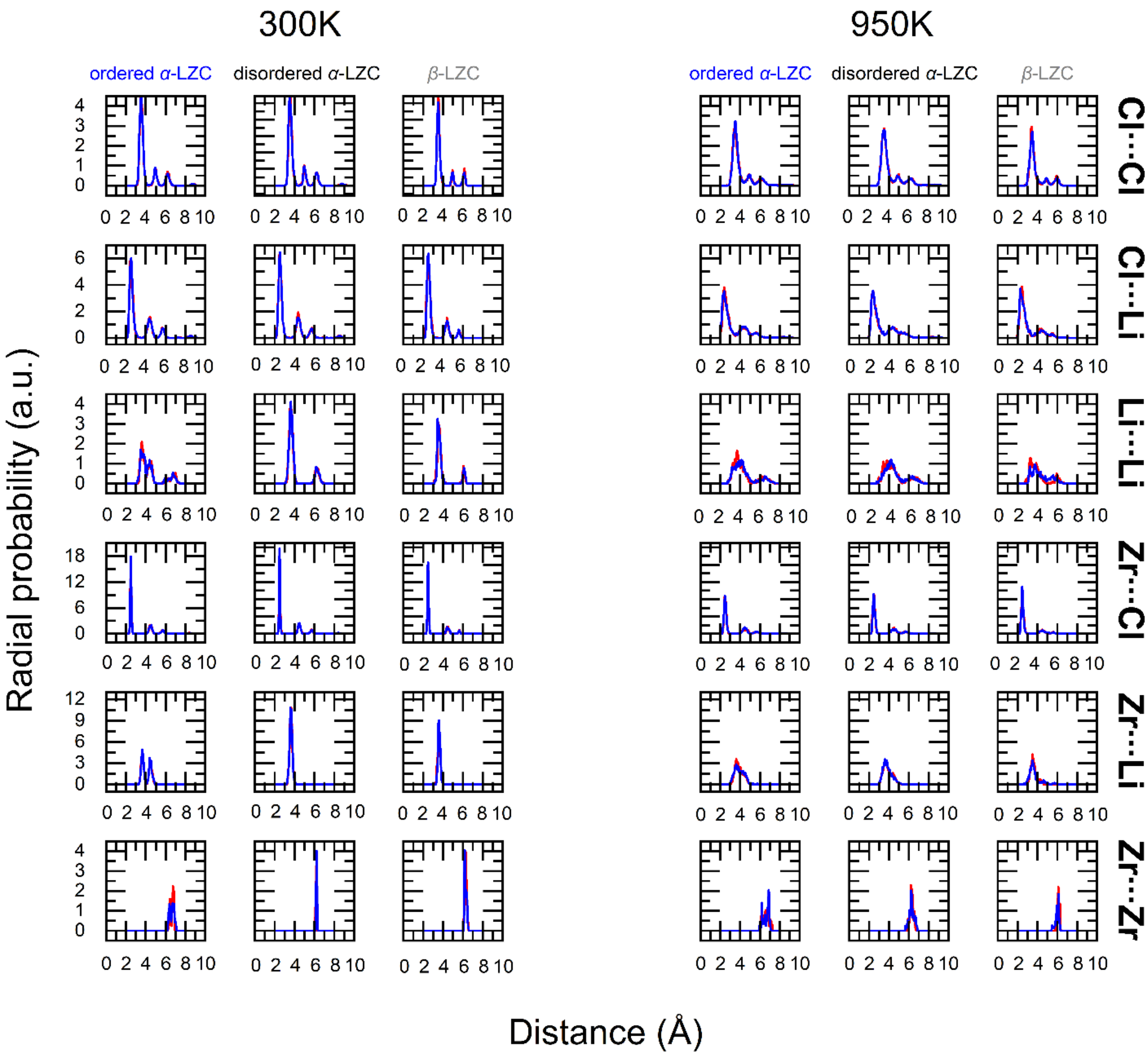

**Figure S4.** Structure characteristics in $Li_2ZrCl_6$, LZC, for ordered *α*-LZC (*left column*), disordered *α*-LZC (*middle column*) and *β*-LZC (*right column*) phases, expressed by the radial distribution functions, RDF, comparison between *red* – *ab initio* molecular dynamics, AIMD, and *blue* – deep learning-accelerated molecular dynamics, DLMD. Comparisons are set at 300 K (*left half*) and 950 K (*right half*), following a 1 ps equilibration period. The results demonstrate that the RDF peak positions, shapes, and heights from DLMD closely match those from AIMD for both room and high temperature, even at extended distances. This high degree of consistency indicates that the deep learning potential model is capable of accurately predicting the static properties of the system at different temperature.

**Table S3.** Lithium-ion transport characteristics in $Li_2ZrCl_6$, LZC, expressed by collective and individual ionic conductivities, $\sigma$ (S cm$^{-1}$), for ordered and disordered $\alpha$-LZC and $\beta$-LZC at different high temperatures by means of deep learning-accelerated molecular dynamics.

| Temperature | collective $\sigma$ | individual $\sigma$ |
|---|---|---|
| | ordered $\alpha$-LZC | |
| 750 K | 0.516 | 0.294 |
| 800 K | 0.710 | 0.395 |
| 850 K | 0.887 | 0.494 |
| 900 K | 0.967 | 0.609 |
| 950 K | 1.201 | 0.716 |
| 1000 K | 1.450 | 0.833 |
| | disordered $\alpha$-LZC | |
| 750 K | 0.610 | 0.274 |
| 800 K | 0.709 | 0.376 |
| 850 K | 0.940 | 0.484 |
| 900 K | 1.053 | 0.585 |
| 950 K | 1.242 | 0.707 |
| 1000 K | 1.463 | 0.823 |
| | $\beta$-LZC | |
| 900 K | 0.280 | 0.232 |
| 950 K | 0.296 | 0.290 |
| 1000 K | 0.360 | 0.352 |
| 1050 K | 0.448 | 0.427 |
| 1100 K | 0.653 | 0.543 |
| 1150 K | 0.690 | 0.619 |

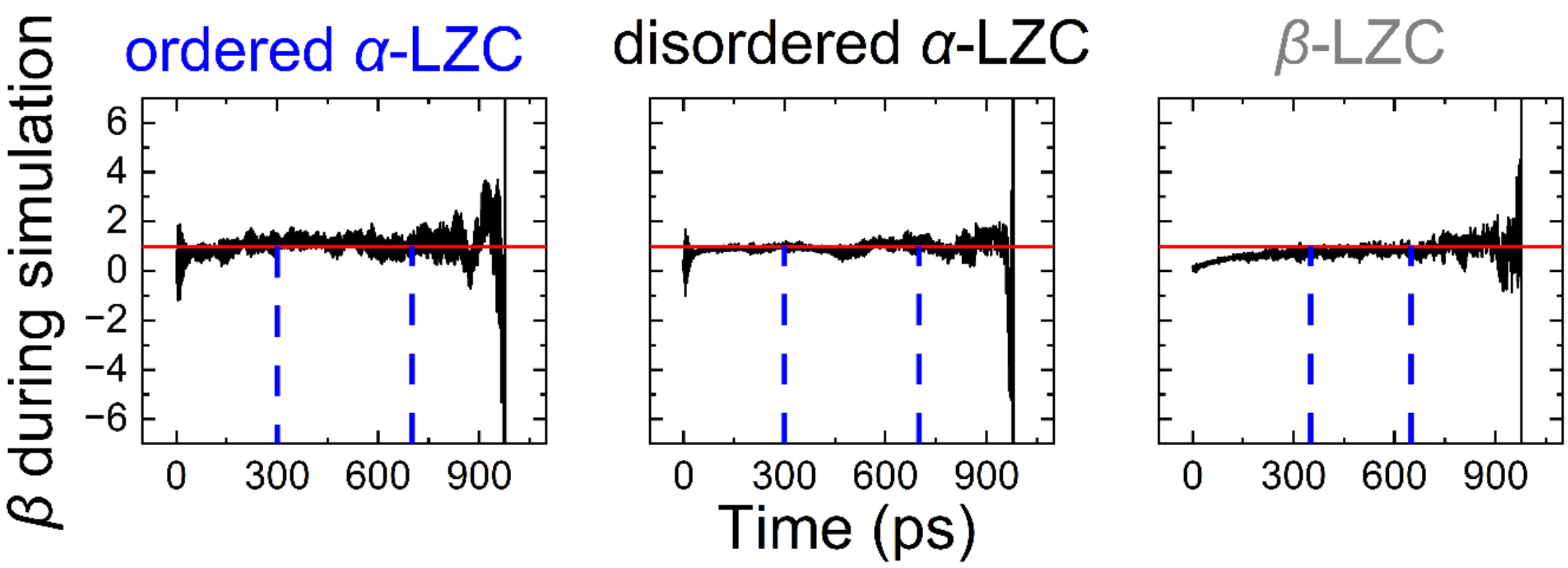

**Figure S5.** Estimated time window for diffusive regime in $Li_2ZrCl_6$, LZC, for ordered $\alpha$-LZC (*left column*), disordered $\alpha$-LZC (*middle column*) and $\beta$-LZC (*right column*) phases at 950K. The *red lines*, $\beta = 1$, indicate Fickian diffusion behavior. Blue dashed lines stand for visual guidance and represent the time window where diffusion regime occurs.

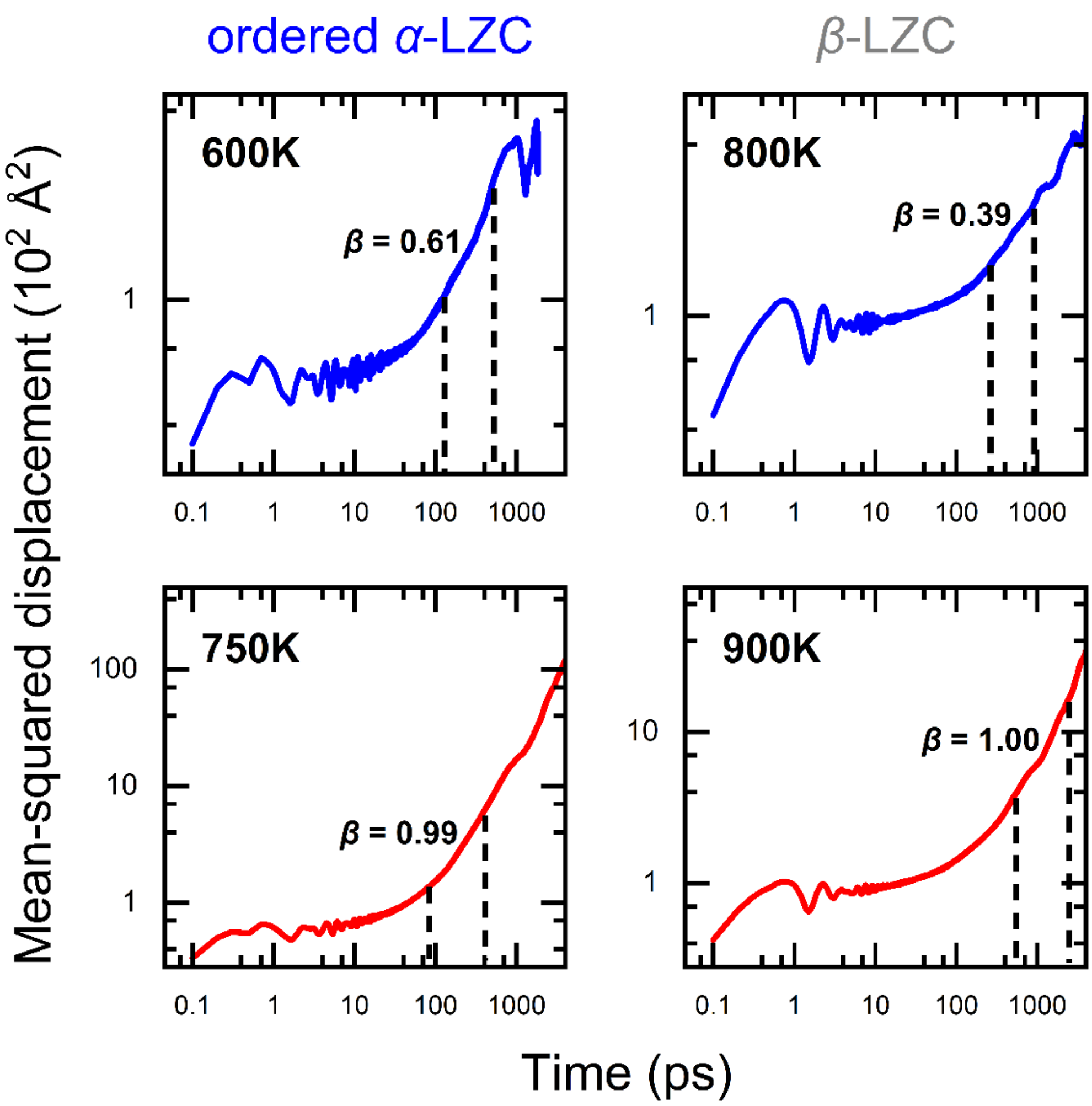

**Figure S6.** Lithium-ion transport characteristics under different temperature in $Li_2ZrCl_6$, LZC, for ordered $\alpha$-LZC (left column) and $\beta$-LZC (right column) phases. Black dashed lines stand for visual guidance and represent the time window where diffusion regime occurs or diffusion coefficient ($\beta$) when no diffusion occurs. The *top raw* represents the temperature examples that cannot reach the diffusion state, which are shown in *blue*. The *bottom raw* represents the temperature examples that can reach the diffusion state, which are shown in *red*, also indicates proper temperature choice for mean-squared displacement.

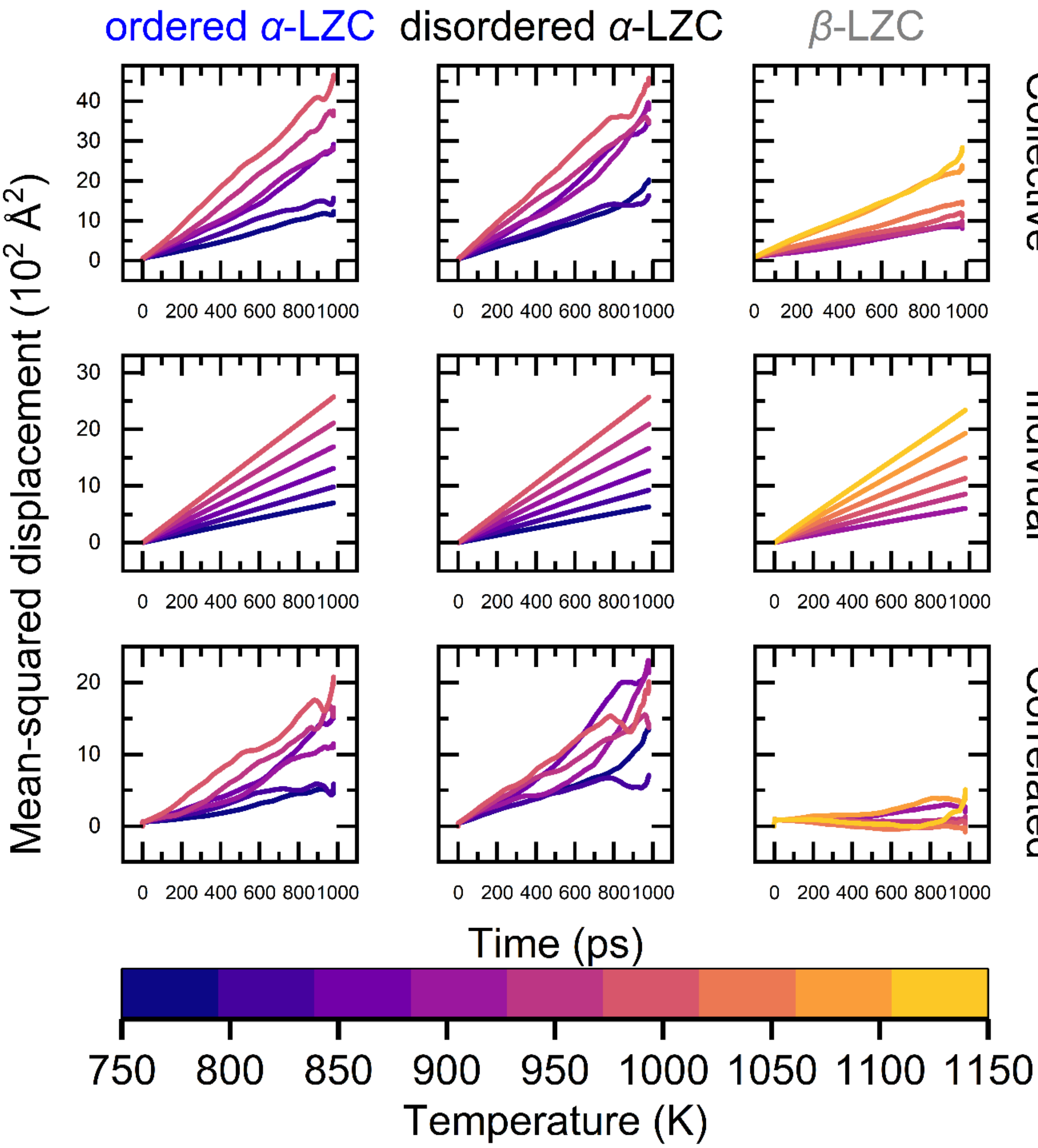

**Figure S7.** Lithium-ion transport characteristics in $Li_2ZrCl_6$, LZC, for ordered $\alpha$-LZC (*left column*), disordered $\alpha$-LZC (*middle column*) and $\beta$-LZC (*right column*) phases, expressed by collective (*top row*), individual (*middle row*) and correlated (*bottom row*) mean-squared displacement of lithium ions obtained by the averaging over 20 runs in temperature range between 750 and 1000 K for $\alpha$-LZC and between 900 and 1150 K for $\beta$-LZC. Both axes are plotted in linear scales. The non-smoothness of correlated subpart indicates its large noise and measurement difficulty. Especially for $\beta$-LZC, the presence of extremely high noise levels suggests a smaller correlated subpart. This is because, under the same noise conditions, a higher noise ratio indicates a reduced contribution from the correlated subpart.

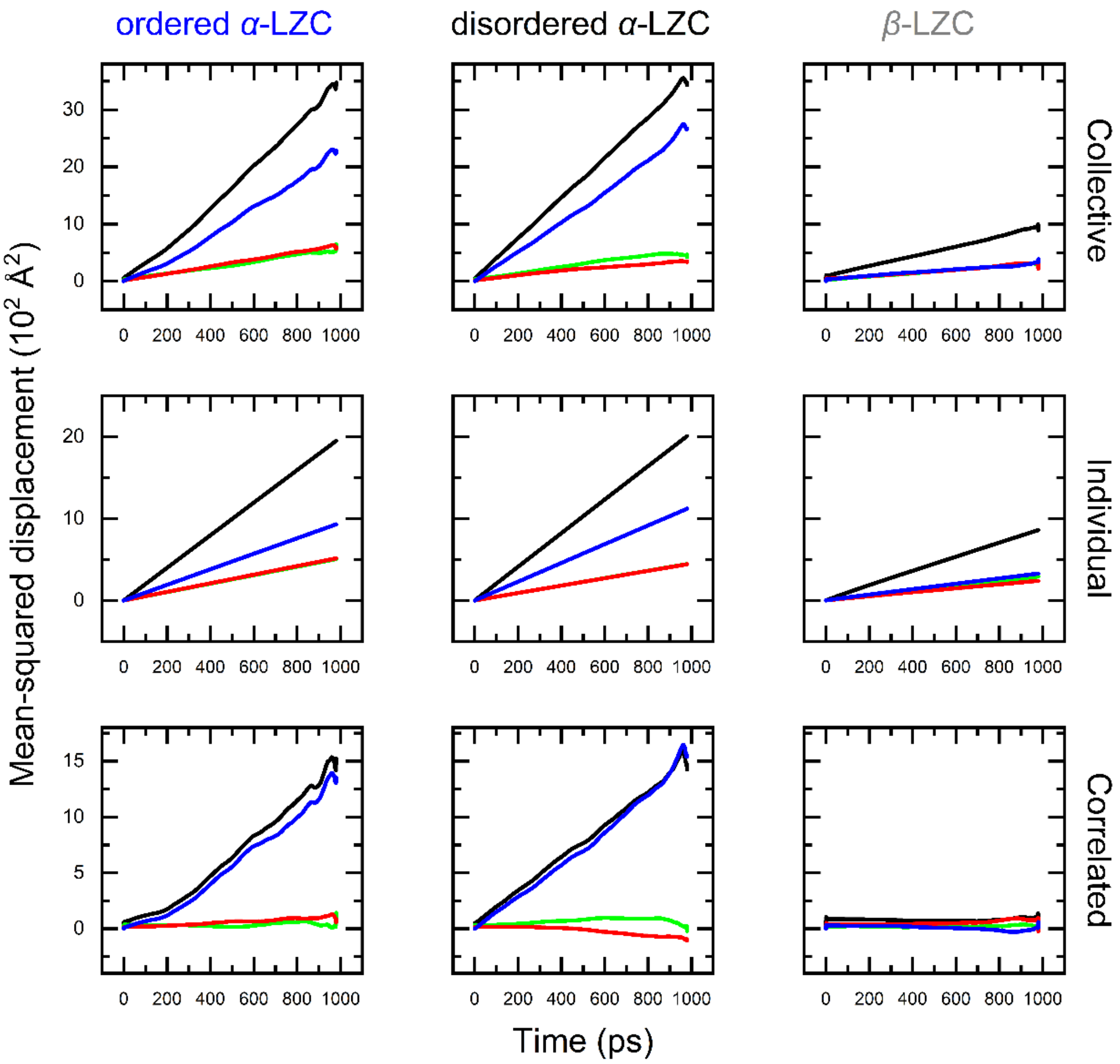

**Figure S8.** Lithium-ion transport characteristics in $Li_2ZrCl_6$, LZC, for ordered $\alpha$-LZC (*left column*), disordered $\alpha$-LZC (*middle column*) and $\beta$-LZC (*right column*) phases, expressed by collective (*top row*), individual (*middle row*) and correlated (*bottom row*) mean-squared displacement of lithium ions with cumulative (*black line*) and directional, *xyz*, terms calculated at 950 K. *Green lines* represent displacement in *x*-direction *red* – *y*-direction, *blue* – *z*-direction. Both axes are plotted in linear scales.

**Table S4.** Lithium-ion transport characteristics in $Li_2ZrCl_6$, LZC, expressed by cumulative, *total*, and directional, *xyz*, individual ionic conductivities, $\sigma$ (S $cm^{-1}$), for ordered and disordered $\alpha$-LZC and $\beta$-LZC calculated at 950 K by means of deep learning-accelerated molecular dynamics.

| | individual $\sigma_x$ | individual $\sigma_y$ | individual $\sigma_z$ | individual $\sigma_{total}$ |
|---|---|---|---|---|
| ordered $\alpha$-LZC | 0.171 | 0.175 | 0.315 | 0.661 |
| disordered $\alpha$-LZC | 0.151 | 0.150 | 0.381 | 0.682 |
| $\beta$-LZC | 0.095 | 0.078 | 0.104 | 0.280 |

**Table S5.** Haven's ratio $H_R$ across different directions to $Li_2ZrCl_6$, LZC, for ordered and disordered $\alpha$-LZC and $\beta$-LZC calculated at 950 K by means of deep learning-accelerated molecular dynamics.

| | $H_{R_x}$ | $H_{R_y}$ | $H_{R_z}$ | $H_{R_{total}}$ |
|---|---|---|---|---|
| ordered $\alpha$-LZC | 0.746 | 0.992 | 0.451 | 0.569 |
| disordered $\alpha$-LZC | 0.933 | 0.833 | 0.553 | 0.687 |
| $\beta$-LZC | 0.896 | 0.785 | 1.057 | 0.920 |

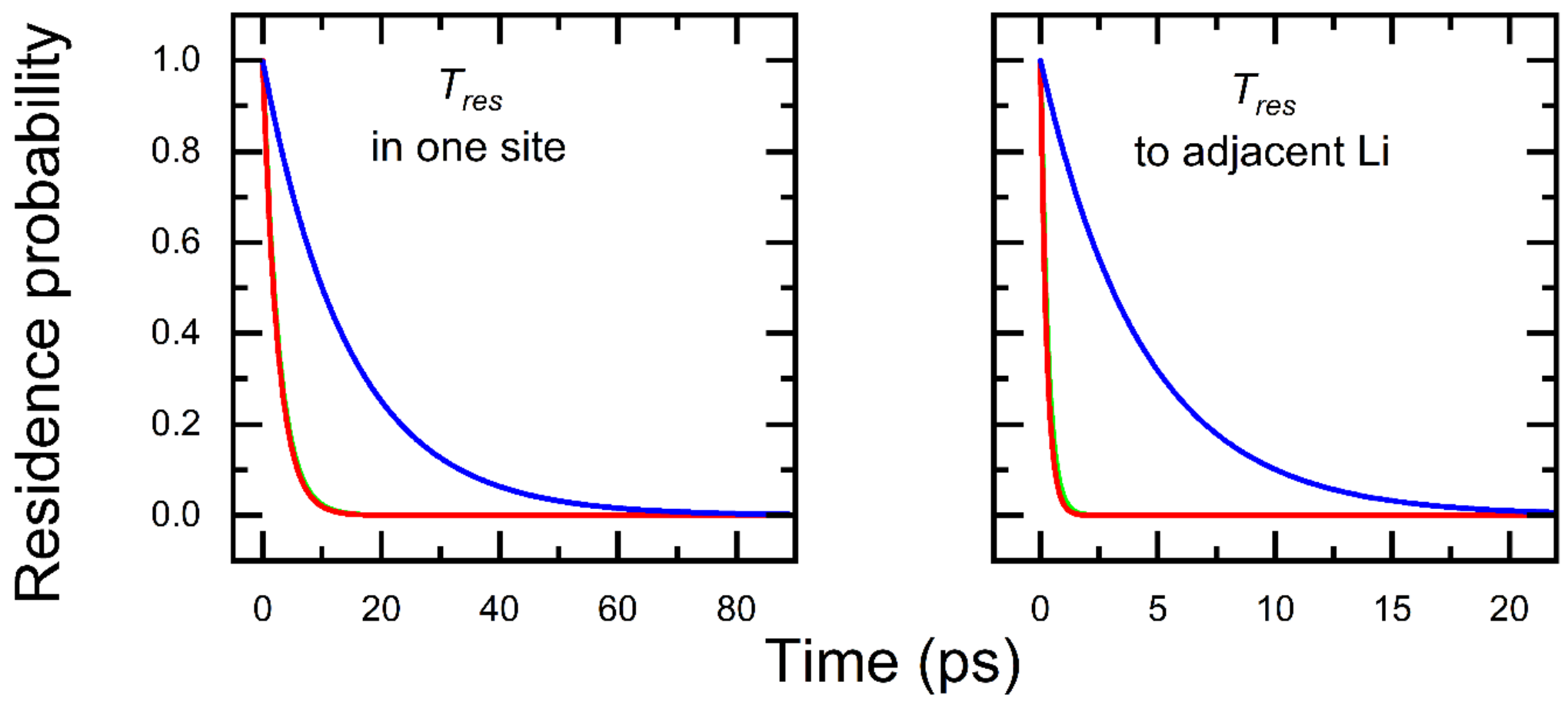

**Figure S9.** Lithium-ion residence probability against time in $Li_2ZrCl_6$, LZC, expressed by the average residence time, $T_{res}$ (ps) in one site (*left column*) and to adjacent lithium (*right column*). When the residence probability approaches 0, that time represents $T_{res}$ which means lithium ion leaves certain dwell state. *Green line* represent ordered $\alpha$-LZC, *red* – disordered $\alpha$-LZC, *blue* – $\beta$-LZC.

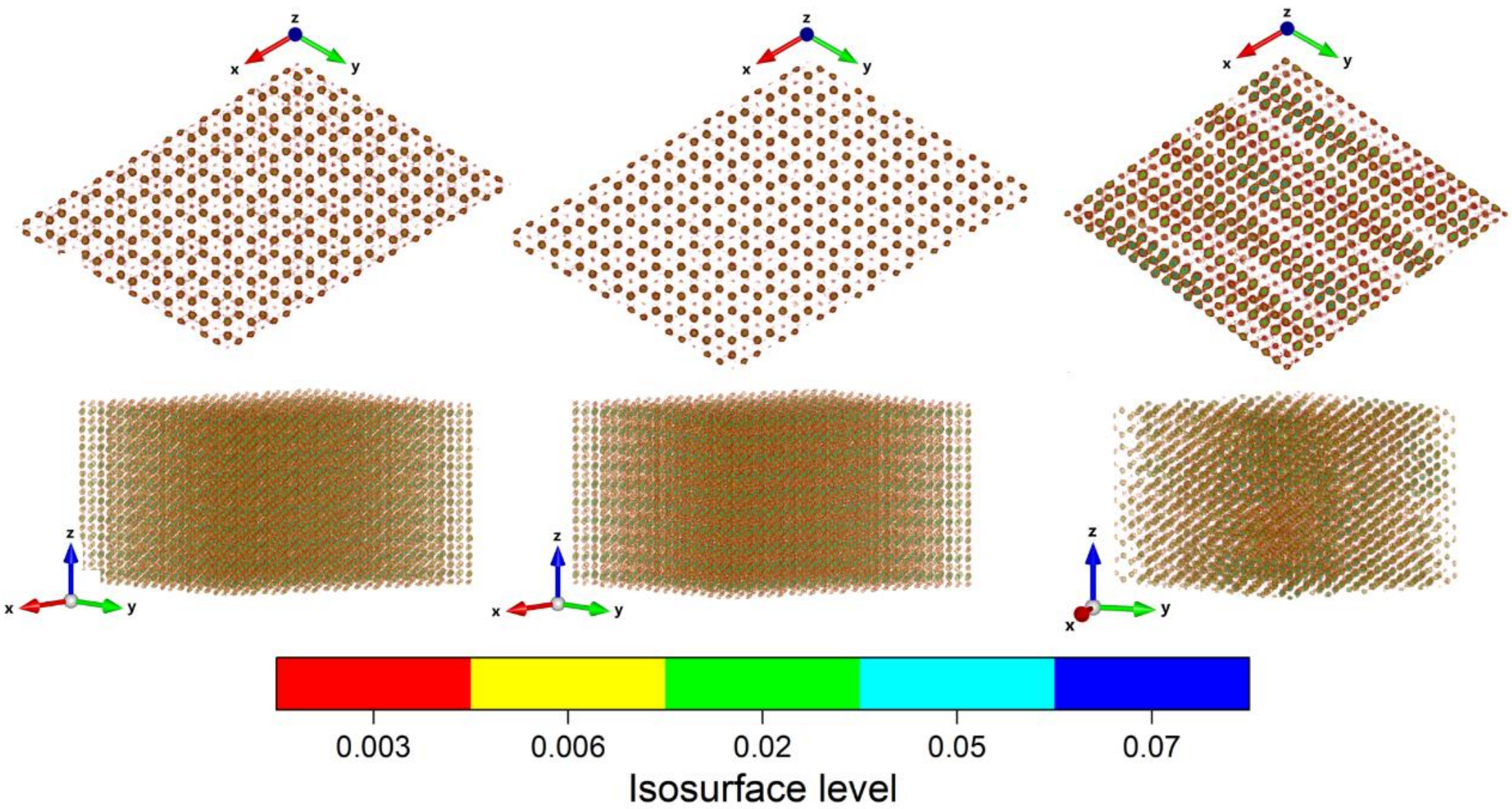

**Figure S10.** Lithium-ion conduction pathway based on deep learning-accelerated molecular dynamics trajectories in $Li_2ZrCl_6$, LZC, for spatial distribution function of lithium-ion in ordered $\alpha$-LZC (*left column*), disordered $\alpha$-LZC (*middle column*) and $\beta$-LZC (*right column*) phases, expressed by top (*upper panel*) and side (*lower panel*) views. Isosurface level unit is Å$^{-3}$.

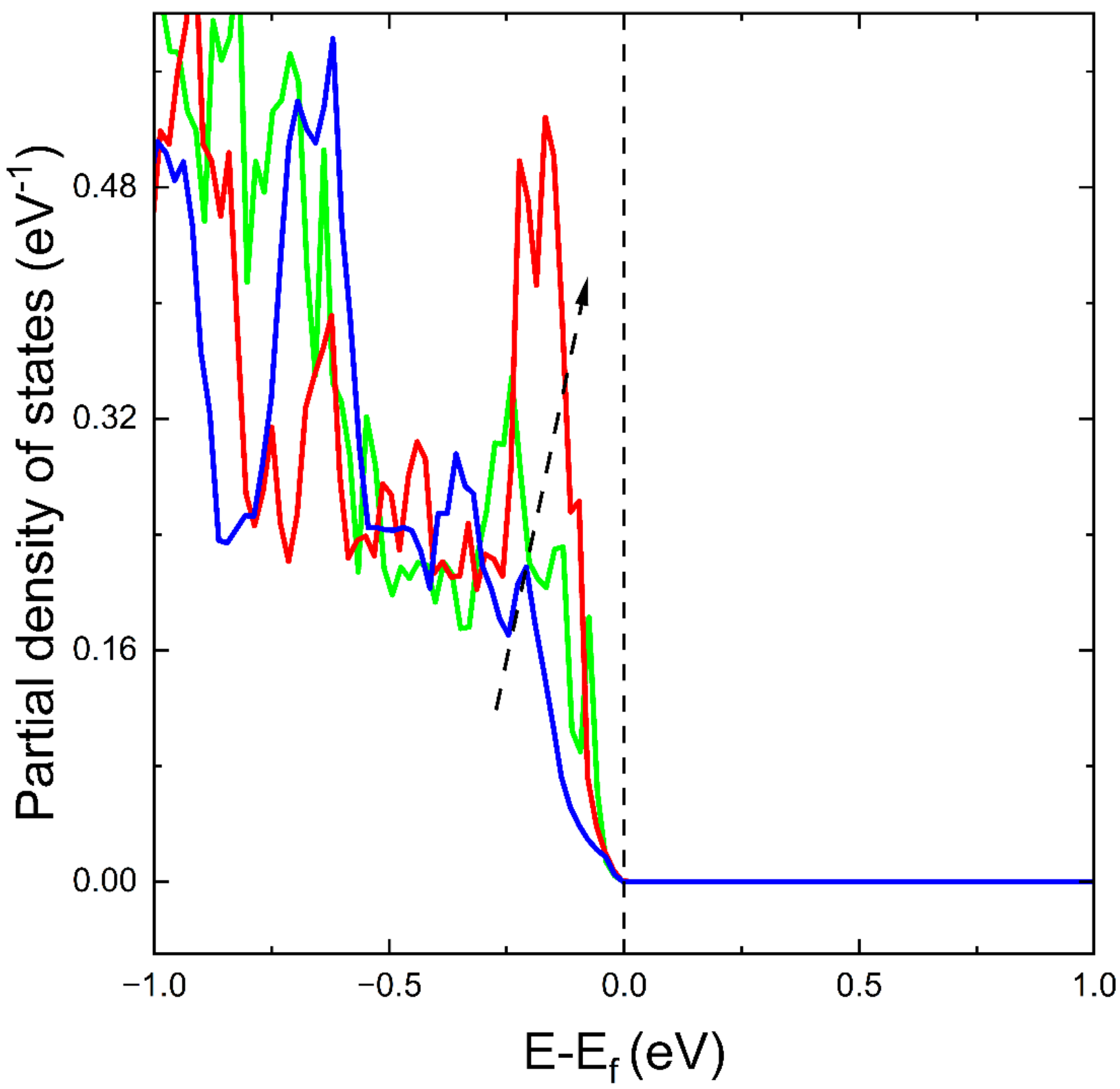

**Figure S11.** Partial density of states of zirconium in $Li_2ZrCl_6$, LZC. The *black dashed line* stands for visual guidance of Fermi level. *Green line* represents ordered *α*-LZC, *red* – disordered *α*-LZC, *blue* – *β*-LZC. The *black dashed arrow* stands for visual guidance of the difference in zirconium occupied state density near the Fermi level. From large to small: disordered *α*-LZC > ordered *α*-LZC > *β*-LZC.

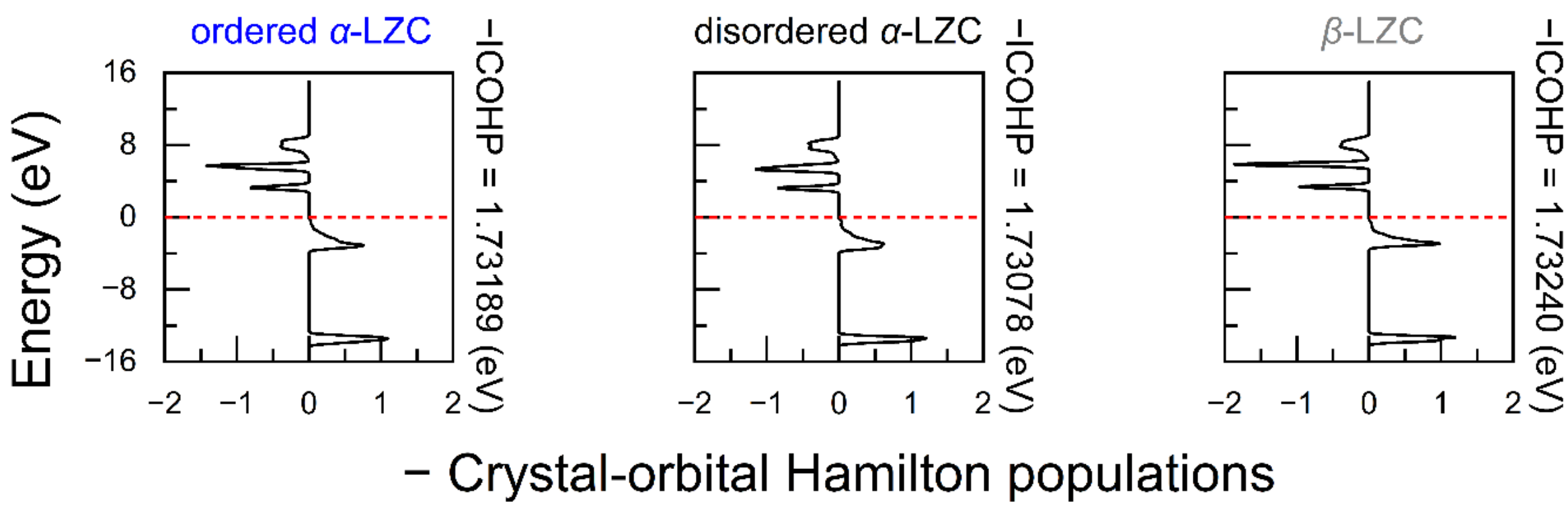

**Figure S12.** Crystal-orbital Hamilton populations of zirconium–chloride bonds in $Li_2ZrCl_6$, LZC, for ordered *α*-LZC (*left column*), disordered *α*-LZC (*middle column*) and *β*-LZC (*right column*) phases. The *red dashed line* stands for visual guidance of Fermi level. The antibonding orbitals are above the Fermi level, and the bonding orbitals are below the Fermi level. Negative integrated crystal orbital Hamilton is marked as −ICOHP on the right of each phase. Larger −ICOHP means stronger zirconium–chloride bonds.